\documentclass[11pt,a4paper]{article}

\usepackage{amssymb,amsmath,amsfonts}
\usepackage{pgf,tikz}
       \usepackage[margin=3cm]{geometry} 
\usepackage{natbib}
\usepackage{hyperref}
\hypersetup{colorlinks=true,
            citecolor=blue,
            linkcolor=blue}
 \usepackage{lipsum}
\newcommand\blfootnote[1]{%
  \begingroup
  \renewcommand\thefootnote{}\footnote{#1}%
  \addtocounter{footnote}{-1}%
  \endgroup
}
\newcommand{\figpath}{./art2/}

\newcommand{\RR}{{\mathbb R}}

\newcommand\vS{{S} }
\newcommand\vK{{K} }
\newcommand\vM{{M} }

\newcommand{\be}{\begin{equation}}
\newcommand{\ee}{\end{equation}}

 {\everymath{\displaystyle\everymath{}}\array}%
 {\endarray}

\begin{document}

\title{Detectability of quasi-circular co-orbital planets. Application to the radial velocity technique.}

\author{A. Leleu\footnote{IMCCE, Observatoire de Paris - PSL Research University, UPMC Univ. Paris 06, Univ. Lille 1, CNRS, 77 Avenue Denfert-Rochereau, 75014 Paris, France} \and  P. Robutel$^*$ \and  A.C.M. Correia$^*$\footnote{Departemento de F\`isica, I3N, Universidade de Aviero, Campus de Santiago, 2810-193 Aveiro - Portugal}
}

\maketitle

 \abstract{Several\blfootnote{E-mail: adrien.leleu@obspm.fr; philippe.robutel@obspm.fr  and correia@ua.pt} celestial bodies in co-orbital configurations exist in the solar system. However, co-orbital exoplanets have not yet been discovered. This lack may result from a degeneracy between the signal induced by co-orbital planets and other orbital configurations. Here we determine a criterion for the detectability of quasi-circular co-orbital planets and develop a demodulation method to bring out their signature from the observational data.  
We show that the precision required to identify a pair of co-orbital planets depends only on the libration amplitude and on the planet's mass ratio.
We apply our method to synthetic radial velocity data, and show that for tadpole orbits we are able to determine the inclination of the system to the line of sight. 
Our method is also valid for planets detected through the transit and astrometry techniques.
}

%

\section{Introduction}

\label{sec:intro}

\citet{Lagrange1772}  found an equilibrium configuration for the three-body problem where the bodies are located at the vertices of an equilateral triangle. For relatively small eccentricities, the libration around the stable Lagrangian equilibrium points $L_4$ and $L_5$ is one of the two possible configurations of a stable co-orbital system, called a tadpole orbit (by analogy with the restricted three-body problem, we define $L_4$ as the equilibrium point when the less massive planet is $60^\circ$ ahead of the more massive one and $L_5$ when it is behind). 
The first object of this kind was observed by \cite{wolf1906}, the asteroid Achilles, which shares its orbit with Jupiter around $L_4$. At present, more than 6000 bodies in tadpole orbits are known in the solar system \citep{MPC}. For objects in the second configuration, called a horseshoe orbit after the shape the trajectories of the bodies in the corotating frame, the libration encompasses the equilibrium points $L_4$, $L_5$, and $L_3$.
A single example is known, for a pair of satellites of Saturn \citep[see][]{DeMu1981b}. 

The Lagrangian equilibria points are stable if the masses of the planets are low enough. In the quasi-circular case, \citet{Ga1843} showed that there is a stability condition for the Lagrangian equilibrium
\be
\frac{m_0 m_1+m_1 m_2+m_0 m_2} {(m_0+m_1+m_2)^2} < \frac{1}{27} \approx 0.037 \ ,
\label{gascheau}
\ee
where $m_0$ is the mass of the star, and $m_1$ and $m_2$ the mass of the co-orbital planets. The mass repartition between the two co-orbitals has a small impact on the stability. Within this limit, Gascheau's criterion guarantees the stability of the linearized equations in the vicinity of $L_4$ or $L_5$. The lower the masses of the co-orbitals with respect to the total mass, the larger is the possible libration amplitude. The horseshoe domain is stable when the planets have a Saturn-mass or less \citep{LauCha2002}. 

For eccentric orbits, the range of stable mass ratios between the co-orbital and the central body decreases as the eccentricity increases \citep{2000Roberts, Nauenberg2002}. Moreover, an additional co-orbital configuration exists in the eccentric case, called quasi-satellite, as the co-orbitals seem to gravitate around each other in the rotating frame. For high eccentricities, co-orbitals have a much larger stable domain for quasi-satellites than for tadpole or horseshoe configurations \citep{GiuBeMiFe2010}.

Since the discovery of the first exoplanets \citep{WoFr1992}, a great diversity of systems has been found, some of them in mean motion resonances (MMR). A few of these resonant systems are highly populated (like the $2/1$ MMR), but so far no system has been identified in a co-orbital configuration ($1/1$ MMR). However, many theoretical works suggest that co-orbital exoplanets may also exist. \citet{LauCha2002} introduced two possible processes that form these systems: (i) planet-planet gravitational scattering and (ii) accretion \textit{\emph{in situ}} at the L4-L5 points of a primary. 
 
The assumptions made on the gas disc density profile in scenario (i) can either lead to systems with a high diversity of mass ratio \citep{CreNe2008} or to equal mass co-orbitals when a density jump is present \citep{GiuBe2012}. In their model, \citet{CreNe2008} form co-orbitals in over $30\%$ of the runs. Initially in horseshoe configurations, the co-orbital systems are generally damped into tadpole configurations. The co-orbitals formed by this process usually have very low inclinations and eccentricities ($e<0.02$).

\citet{LyJo2009} showed that in scenario (ii), up to 5-20 Earth-mass planets may form in the tadpole of a Jupiter-mass primary. 
For existing co-orbitals, the gas accretion seems to increase the mass difference between co-orbitals, the more massive of the two reaching Jovian mass while the starving one stays below $70 M_\oplus$ \citep{CreNe2009}.

 Models that produce co-orbital planets due to dissipation within a disc may also experience significant inward migration. Such migration increases the libration amplitude of trojan planets for late migrating stages with low gas friction. This may lead to instability, but the damping of the disc usually forces the libration to remain small. Equal mass co-orbitals (from super-Earths to Saturns) are heavily disturbed during large scale orbital migration \citep{PiRa2014}. In some cases, \citet{RoGiMi2013} have shown as well that long-lasting tidal evolution may perturb equal mass close-in systems. 
Overall, significantly different mass trojans may thus be more common, especially in close-in configurations.

Co-orbital planets can also be disturbed in the presence of additional planetary companions, in particular by a significantly massive planet in another MMR with the co-orbitals \citep{MoLeTsiGo2005, RoBo2009}. Moreover, horseshoe orbits are more easily disturbed than tadpole orbits owing to the higher variation of  frequencies in the system with respect to the libration amplitude \citep{RoPo2013}.

Detecting co-orbital planets is not an easy task because the signatures of each body are usually superimposed. 
The transit method can solve this degeneracy, either by observing both co-orbitals transiting \citep{Ja2013}, or when only one co-orbital is transiting by coupling with another detection method \citep{FoGa2006}.
When the libration amplitude is large enough, the transit timing variation (TTV) method can also find co-orbital candidates even when only one body is transiting \citep{Ja2013, VoNe2014}.

If the bodies are exactly at the Lagrangian equilibrium point (with no libration), the radial velocity (RV) signal is the same as for a single planet, overestimating the mass of the system by $\approx 13\%$ \citep{Dobrovolskis2013}. \citet{LauCha2002} showed that the libration of co-orbitals modulates the RV signal from the star, allowing them to determine a co-orbital system from a simulated RV signal. \citet{CreNe2009} estimated that for the co-orbitals obtained in their simulation, this modulation is  $10\% - 20\%$ of the total amplitude of the signal. \citet{GiuBe2012} showed that a short-term RV signal (duration inferior to the period of libration) did not allow  co-orbitals to be distinguished from an eccentric planet or from two planets in $2/1$ MMR \citep{GoKo2006, AELM2010}.

Theoretical and numerical studies seem to agree that tadpole and horseshoe co-orbitals tend to have low eccentricities and mutual inclinations. In addition, in the solar system we observe among the moons of Saturn \citep[][and references therein]{DeMu1981b} three co-orbital systems, all with inclinations $<1^\circ$ and eccentricities  $<0.01$. Jupiter's trojans have a mean inclination of about $\approx 12^\circ$ and eccentricities bellow $0.3$. We thus conclude that quasi-circular orbits are a good approximation for many co-orbital systems. As the study of quasi-circular and coplanar orbits is easier than the full case, in this paper we restrict our analysis to this  simpler situation.
We first give a brief overview of the co-orbital dynamics and an analytical approximation of the co-orbital motion valid for small eccentricities. In section 3 we derive an analytical method that can be used to extract the signature of co-orbitals from the motion of the host star. In section 4 we exemplify our method with the radial velocity technique using synthetic data, and we conclude in the last section.

\section{Co-orbital dynamics}
\label{sec:coo}

\begin{figure}[h!]
\begin{center}
\includegraphics[width=.6\linewidth]{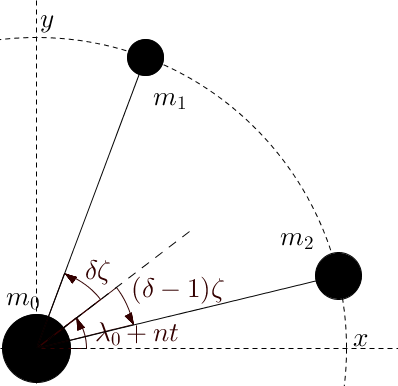}
\caption{\label{fig:angles} Reference angles represented for the circular co-orbital system with respect to an inertial frame $(x,y)$. $m_0$ is the mass of the central star, $m_1$ and $m_2$ the masses of the co-orbitals. $r_i$ is the distance of the co-orbital $i$ to the central star, and $\lambda_i$ its true longitude. Following equation (\ref{eq:lamb}) one can write $\lambda_i$ as a function of $\lambda_0$, $\zeta$, and of the mass ratio $\delta$.}
\end{center}
\label{fig:angles}
\end{figure}

We denote $m_0$ the mass of the central star, and $m_1$ and $m_2$ the masses of the co-orbital planets. Adapting the theory developed by \'Erdi (1977) for the restricted three-body problem to the case of three massive bodies, and neglecting the quantities of second order and more in 
\be
\mu = \frac{m_1+m_2}{m_0 + m_1+m_2} \ ,
\ee
the mean longitudes $\lambda_i$ and the semi-major axes $a_i$ of the co-orbitals can be approximated by the expressions \citep[see][]{RoRaCa2011} 
\begin{equation}
 \begin{array}{ll}
\lambda_1(t) \approx   nt + \delta \zeta(t) + \lambda_0 \ , & \lambda_2(t)\approx  nt - (1-\delta) \zeta(t) + \lambda_0 \ , \\ \\
a_1(t) \approx   a \left(1 -\delta \frac23 \frac{\dot\zeta(t)}{n} \right), &  a_2(t) \approx   a \left(1 + (1-\delta)\frac23 \frac{\dot\zeta(t)}{n} \right) \ ,
\end{array}
\label{eq:lamb} 
\end{equation}
where $n$ is the averaged mean motion of the barycentre of $m_1$ and $m_2$, 
$\lambda_{0}$ is a constant equal to the initial value of the barycentre of the longitudes $(\lambda_1 m_1+\lambda_2 m_2)/(m_1+m_2)$, and
\be
\delta = \frac{m_2}{m_1 + m_2} \ . 
\ee
At  first order in $\mu$, we have the constant quantity $a = (1-\delta)a_1 + \delta a_2$, which can be seen as the mean semi-major axis, and is linked to $n$ by  Kepler's third law $n^2 a^3 = G (m_0 + m_1 + m_2)$, where $G$ is the gravitational constant. Finally, the variable $\zeta = \lambda_1 - \lambda_2$ satisfies the second-order differential equation 

\begin{equation}
\ddot\zeta=-3\mu n^2\left[1-(2-2\cos\zeta)^{-3/2}\right]\sin\zeta \ ,
\label{eq:sol_orb_zeta} 
\end{equation}%
which is one of the most common representations of the co-orbital motion \citep[see][and references therein]{Morais1999}. For small eccentricities, it describes the relative motion of the two bodies  and is valid as long as the co-orbital bodies are not too close to the collision ($\zeta=0$).  

\begin{figure}[h!]
\centering
 \includegraphics[width=.8\linewidth]{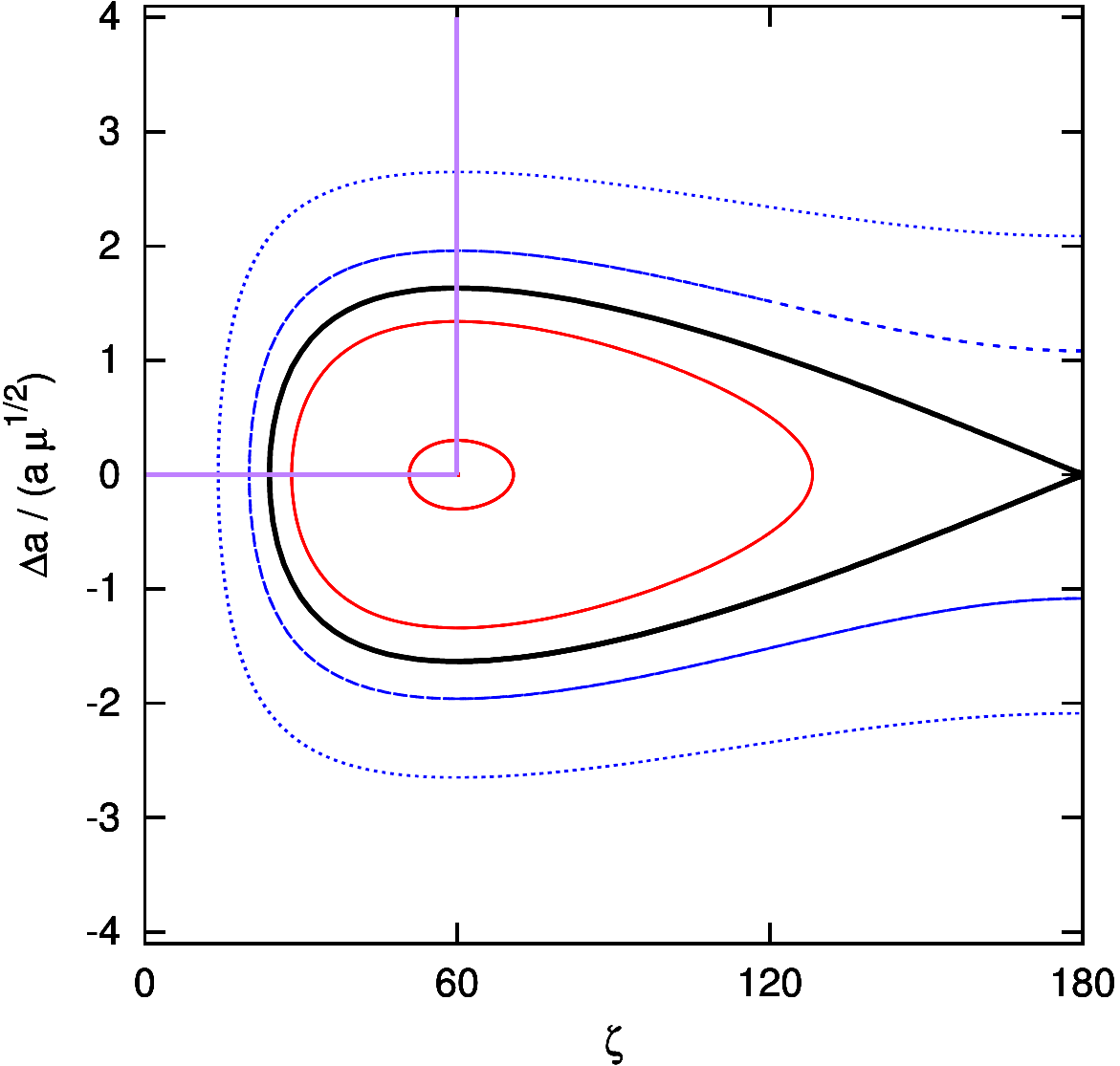}
\caption{Phase portrait  of  equation (\ref{eq:sol_orb_zeta}). The separatrix (black curve) splits the phase space in two different domains: inside the separatrix the region associated with the tadpole orbits (in red) and the horseshoe domain (blue orbits) outside. The phase portrait is symmetric with respect to $\zeta=180^\circ$.
The horizontal purple segment indicates the range of variation of $\zeta_0$ while the vertical one shows the section used as initial condition to draw Fig. \ref{fig:stabzp}. See the text for more details.
}
\label{fig:orb}
\end{figure}

Equation (\ref{eq:sol_orb_zeta}) is invariant under the symmetry $\zeta\longmapsto 2\pi-\zeta$, so the study of its phase portrait can be reduced to the domain $(\zeta,\dot\zeta) \in [0, \pi]\times \RR$  (see Fig. \ref{fig:orb}) (this symmetry will be developed later on). The equilibrium point located at $(\zeta,\dot\zeta) = (\pi/3,0)$ corresponds  to one of the two Lagrangian equilateral configurations\footnote{The coordinates of the other equilateral point are $(5\pi/3,0)$. The permutation of the index $1$ and $2$ of the planets allows  the two equilateral configurations to be exchanged.}, which are linearly stable for sufficiently small planetary masses (see below).
Another equilibrium point, whose coordinates are $(\pi,0)$, corresponds to the unstable Eulerian collinear configuration of the type $L_3$. The separatrices emanating from this last unstable point split the phase space in three different regions: two corresponding to the tadpole trajectories surrounding one of the two Lagrangian equilibria (in red in Fig.\,\ref{fig:orb}), and another one corresponding to the horseshoe orbits that surround the three above-mentioned fixed points (in blue in Fig.\,\ref{fig:orb}).

As shown in Figure~\ref{fig:orb}, any trajectory given by a solution of equation (\ref{eq:sol_orb_zeta}) can be entirely determined by the initial conditions $(t_0,\zeta_0)$ such that $\zeta(t_0)=\zeta_0$ and $\dot\zeta(t_0) =0$, where $\zeta_0$ is the minimum value of $\zeta$ along the trajectory, and $t_0$ the first positive instant for which the value $\zeta_0$ is reached. \\
The possible values of $\zeta_0$, represented by the purple horizontal line in Figure \ref{fig:orb}, are included in the interval $]0^\circ,60^\circ]$; $\zeta_0 = 60^\circ$ corresponds to the equilateral configuration where $m_1$ is the leading body and $m_2$ is the trailing one\footnote{The permutation of these two bodies exchanges the two equilateral configurations, which are located at $\zeta = \pi/3$ and  $\zeta = 5\pi/3$, respectively.}. The tadpole orbits are associated with $\zeta_0 \in ]\zeta_s, 60^\circ[$, $\zeta_s \approx 23.9^\circ$  being associated with the separatrix, while $\zeta_0$ ranges from $\zeta_s$ to $0$ for horseshoe orbits.
As a result, the shape of the trajectory of the  relative motion (as the libration amplitude of the resonant angle $\zeta$) is entirely determined by the quantity $\zeta_0$. 

In contrast, $t_0$ and $n\sqrt{\mu}$ are necessary to know the exact position on the trajectory, and in particular, the amplitude of the variations of the semi-major axes. We can rewrite equation (\ref{eq:sol_orb_zeta}) by rescaling the time with $\tau =  \sqrt{\mu}nt$,
\begin{equation}
\frac{d^2 \tilde{\zeta}}{d\tau^2}=-3\left[1-(2-2\cos\tilde{\zeta})^{-3/2}\right]\sin\tilde{\zeta} \ ,
\label{eq:sol_orb_zeta_tau} 
\end{equation}%
where $\tilde{\zeta}(\tau)=\zeta(t)$. As a consequence, this differential equation does not depend on $n\sqrt{\mu}$. Its solutions are solely determined by the initial conditions $\tau_0=\sqrt{\mu}n t_0$ and $\tilde{\zeta}(\tau_0)=\tilde{\zeta}_0 \equiv \zeta_0$.

 \begin{figure}[h!]
\centering
 \includegraphics[width=.9\linewidth]{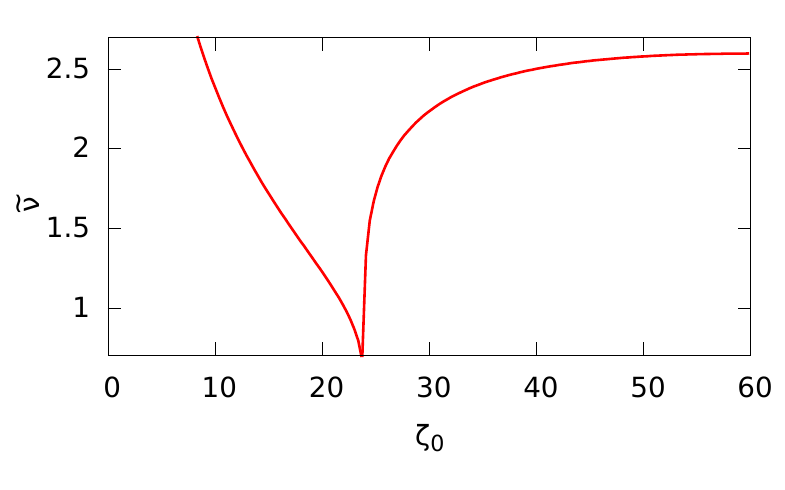}
\caption{
Variation of the libration frequency $\tilde{\nu}$ versus $\zeta_0=\tilde{\zeta}_0$. The frequency is  taken
over the purple horizontal line in Fig. \ref{fig:orb}.
Inside the tadpole region, the libration frequency decreases from $\sqrt{27/4}$ at $L_4$ ($\zeta_0 = 60^\circ$) to $0$ on the separatrix ($\zeta_0 = \zeta_s \approx 23.9^\circ$).
In the horseshoe domain ($\zeta_0<\zeta_s$) the frequency increases from $0$ on the separatrix to infinity when the two planets get closer  because the approximations leading to  Eq. (\ref{eq:sol_orb_zeta}) are not valid close to the collision \citep[see][]{RoPo2013}. 
}
\label{fig:zr}
\end{figure}

In a small vicinity of the Lagrangian equilibria, the frequencies of the motion are close to
\be
\nu_0 = n \sqrt{\frac{27}{4}\mu}.
\label{eq:nu_L}
\ee
More generally, excluding the separatrix, the solutions of equation (\ref{eq:sol_orb_zeta}) (respectively (\ref{eq:sol_orb_zeta_tau})) are periodic. The associated frequency, denoted by $\nu$ (respectively $\tilde{\nu}$), depends on the considered trajectory. However, the time-normalized frequency associated with equation (\ref{eq:sol_orb_zeta_tau}), 
\be
\tilde{\nu}=\nu/(n\sqrt{\mu}) \ ,
\label{eq:nut}
\ee
depends only on $\zeta_0$ ($\tilde{\nu}$ is plotted versus $\zeta_0$ in Fig. \ref{fig:zr}). In tadpole configurations, this dimensionless frequency remains almost constant in the vicinity of the Lagrangian equilibrium $\nu \approx \nu_0$ (Eq.\,\ref{eq:nu_L}) and tends to $0$ as the separatrix is reached at $\zeta_0 = \zeta_s$. In horseshoe configurations, $\nu$ can take any value. In  Figure~\ref{fig:zr}, one can see that far from the separatrix, $\tilde{\nu}$ is always of order unity. This imposes that the variations of the difference of the longitudes, $\zeta$, are slow with respect to the orbital time scale, i.e. $\nu \ll n$. It turns out that $\dot\zeta(t)/n \ll 1$ and as a consequence,  the quantities $a_j$ can be approximated by $a$ (Eq.\,\ref{eq:lamb}). Thus, in the circular planar case, at order $0$ in $\nu$, the position of $m_1$ and $m_2$ in the heliocentric system $\bold{r}=(x + i y)$ are given by
\be
\bold{r_1}=a e^{i \delta \zeta} e^{i(nt + \lambda_0)} \ , \ \text{and}\ \ \bold{r_2}=a e^{- i (1-\delta) \zeta} e^{i(nt + \lambda_0)} \ .
\label{eq:r_v1}
\ee
%
Within the same approximation, we can also write the derivative of previous equation, which gives us the heliocentric velocity of the co-orbitals
\be
\bold{\dot{r}_1}=ina  e^{i \delta \zeta}e^{i(nt + \lambda_0)} \ , \ \text{and}\ \ \bold{\dot{r}}_2=ina  e^{- i (1-\delta) \zeta}e^{i(nt + \lambda_0)} \ . 
\label{eq:rp_v1}
\ee

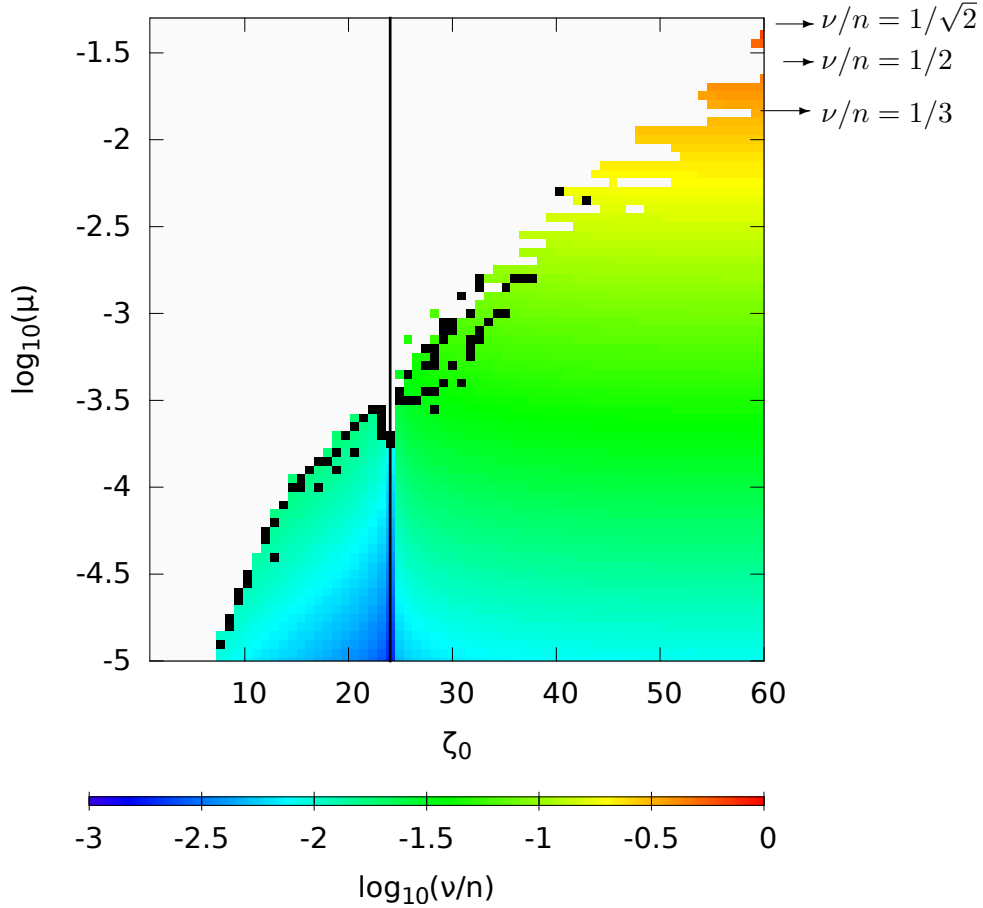
\begin{figure}[h!]
\begin{center}
\includegraphics[width=.8\linewidth]{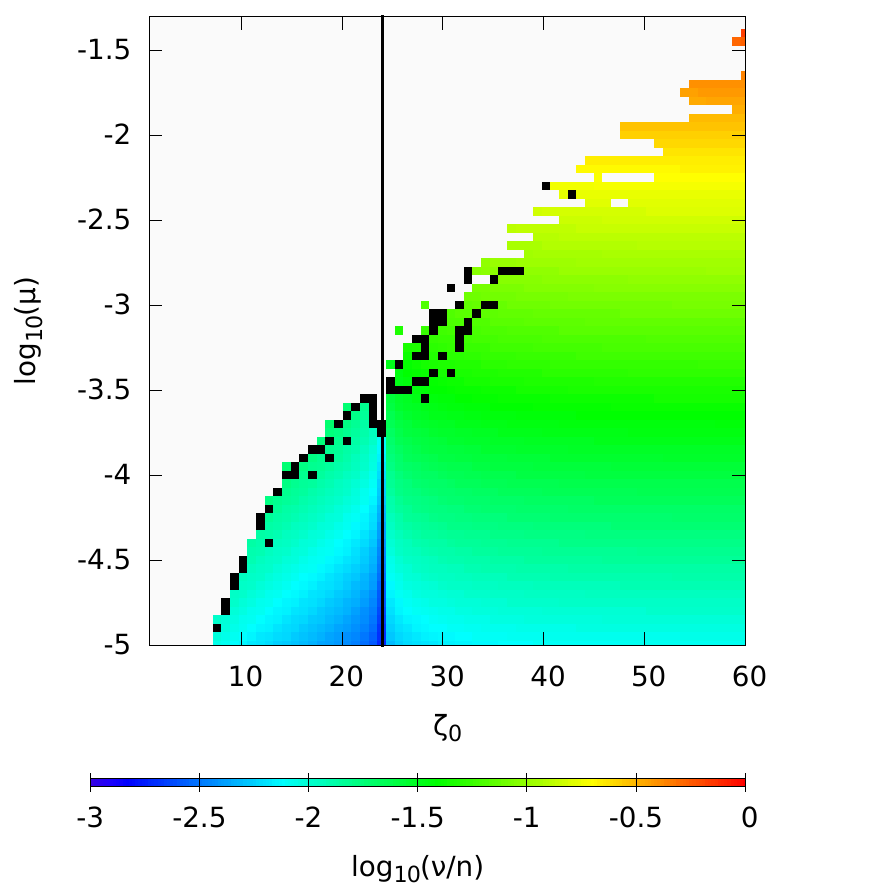}
  \setlength{\unitlength}{1cm}
\begin{picture}(.001,8)
\put(-2,11.7){\vector(1,0){0.5}}
\put(-1.4,11.65){$\nu/n=1/\sqrt{2}$}
\put(-1.9,11.2){\vector(1,0){0.4}}
\put(-1.4,11.1){$\nu/n=1/2$}
\put(-2.2,10.55){\vector(1,0){0.655}}
\put(-1.4,10.4){$\nu/n=1/3$}
\end{picture}

\caption{\label{fig:stabzr} Stability of co-orbitals as a function of $\log_{10}(\mu)$ and $\zeta_0$. The initial conditions are chosen as $t_0=0$ ($\Delta a/a=0$) and $\zeta_0 \in [0^\circ,60^\circ]$: purple horizontal line in Figure \ref{fig:orb}. In black is the separatrix between the tadpole and the horseshoe domain. The stability criteria of \citet{Ga1843}, corresponding to $\nu/n=1/\sqrt{2}$, has been indicated. We also show the vicinity of two of the main resonances between $\nu$ and $n$ : the $1/2$ resonance \citep[see][]{2000Roberts} and the $1/3$. The colour code indicates the value of the libration frequency, i.e. $log_{10}(\nu/n)$.}
\end{center}
\end{figure}

\begin{figure}[h!]
\begin{center}
\includegraphics[width=.8\linewidth]{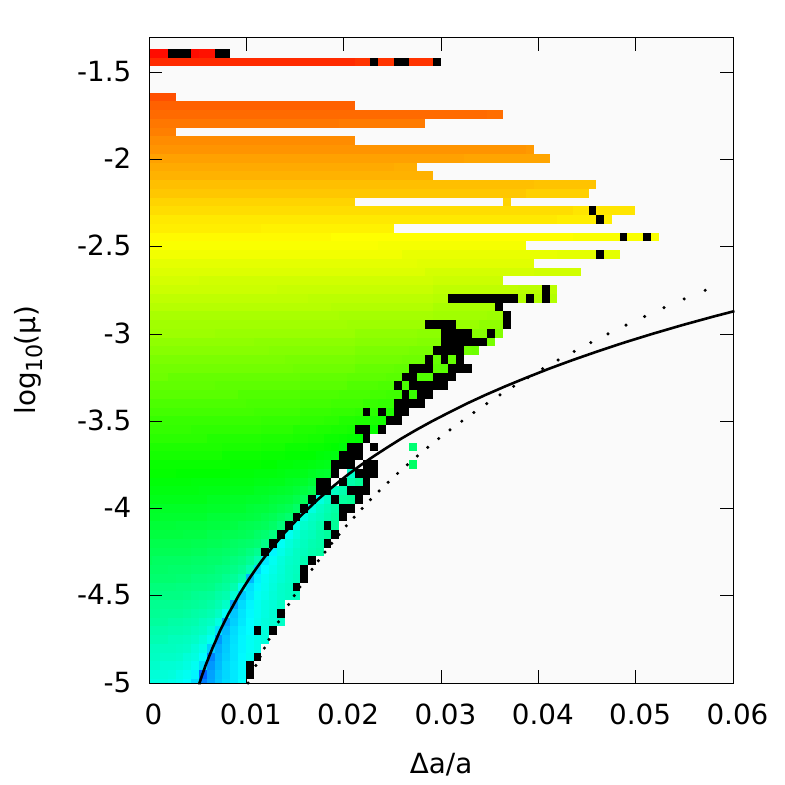}
\caption{\label{fig:stabzp} Stability of co-orbitals as a function of $log_{10}(\mu)$ and $\Delta a/a$. The initial conditions are $\zeta_0=\pi/3$ and $\Delta a/a \in [0,0.06]$: vertical purple line in Figure \ref{fig:orb}. The black line indicates the separatrix between the tadpole and the horseshoe domains. The dots follow a curve $\Delta a \propto \mu^{1/3}$, delimiting the stability region of the horseshoe domain. The colour code indicates the value of the libration frequency, i.e. $log_{10}(\nu/n)$ (see Figure \ref{fig:stabzr} for the scale).}
\end{center}
\end{figure}

%

While searching for co-orbital bodies, the stability of each configuration also needs to be taken into account. 
In order to determine the influence of the planetary masses on the global stability of planar co-orbital systems, we  show the results of two numerical simulations  indicating the width of the stable co-orbital region in different directions.  In Figure \ref{fig:stabzr}, we consider two planets orbiting around a star of mass $m_0=1\,M_\odot$, with fixed initial elements $a_1 = a_2 =1$~au, $e_1 = e_2 =0.05$, and $\lambda_1=\varpi_1=0$, and we vary the initial element $\lambda_2=\varpi_2=-\zeta_0$ in $[0^\circ,60^\circ]$ and their masses $m_1=m_2 = \mu m_0/2$, with $\mu/2 \in [10^{-8}, 10^{-1}]$. For each set of initial conditions, the system is integrated over 5~Myr using the symplectic integrator SABA4 \citep{LaRo2001} with a time-step of $0.0101$~year.

Strongly chaotic systems or systems that quit the co-orbital resonance before the integration stops are removed from the computation. In this case, in Figure~\ref{fig:stabzr} white dots are  assigned to their initial parameters $(\zeta_0,\mu)$. This strong short-term instability is mainly due to the overlapping of low-order secondary resonances \citep{PaEf2015}. After the elimination of these initial conditions, long-term diffusion along secondary resonances may also destabilize the co-orbital systems on a much longer time scale. Measuring the temporal variation of the libration frequency identifies this diffusion \citep{La90,La99}. The black dots indicate a relative variation of over $10^{-6}$ between the first and second half of the $5$~million years integration (to compare with $\approx 10^{-10}$ for the long-term stable configurations). They are  mainly located in the vicinity of the separatrix and near the ejection boundary. In the remaining regions, the small variation of the frequency $\nu$ guarantees, in most cases, the stability for a billion years \citep{RoGa2006}. For long-term stable systems, a colour depending on its libration frequency $\nu$  is assigned to regular resonant co-orbital systems (see the colour code at the bottom of Figure~\ref{fig:stabzr}).
%
%
%
We observe that for large planetary masses, slightly lower than the limit value $\mu\approx 0.037$ \citep{Ga1843}, the stability region is extremely small and strongly perturbed by low-order secondary resonances. 
The chaos generated by the main secondary resonances, namely the $\nu = n/2$, $\nu = n/3$, and $\nu = n/4$ , shrink   the stability region significantly, reducing it to a small region near  the equilateral configuration \citep[see][]{Robe2002,Nauenberg2002}. As $\mu$ decreases, the width of the stable tadpole region increases, and the destabilizing influence of the secondary resonances becomes dominant only on the boundary of the stability region \citep[see][for the restricted problem]{PaEf2015,RoGa2006,ErNa2007}.
When $\mu \approx 3\times 10^{-4} \approx 2 M_{Saturn}/M_{\odot}$, the whole tadpole domain becomes stable, except for a small region around the separatrix ($\zeta_0 = \zeta_s \approx 23.9^{\circ}$).  On the other side of the separatrix, for $\zeta_0<\zeta_s$, stable horseshoe orbits start to appear  \citep[see][]{LauCha2002}. For lower planetary masses,  the size of the horseshoe orbital domain increases as $\mu$ decreases, to reach the outer boundary of the Hill sphere at a distance to the separatrix of the order of $\mu^{1/3}$ \citep[see][]{RoPo2013}.

In Figure \ref{fig:stabzp} we show another section of the co-orbital region. Instead of varying the angle $\zeta_0$, we change the initial value of the difference of the semi-major axes from the equilateral equilibrium $L_4$ towards the outside of the co-orbital region (vertical purple line in Fig. \ref{fig:orb}). More precisely, the initial conditions of the planetary systems are $e_1 = e_2 =0.05$, $\lambda_1=\varpi_1=0$, $\lambda_2=\varpi_2=\pi/3$, and $a_j = 1 - (-1)^j\Delta a$ with $\Delta a \in [0:0.06]$. As they do in figure \ref{fig:stabzr}, the planetary masses vary as $m_1=m_2 = \mu m_0/2$, with $\mu/2 \in [10^{-8}, 10^{-1}]$.  

The tadpole domaine lies above the solid black line corresponding to the equation $ \Delta a = 2\sqrt{2}/\sqrt{3} \sqrt{\mu}$ \citep{RoPo2013}. In this case, contrarily to the $\zeta_0$ direction where the width of the stable tadpole region is a monotonous function of $\mu$, the extent of the stability region reaches a maximum for $\Delta a/a \approx 0.052$  at $\mu = 3.5\times 10^{-3}$ and then tends to zero with $\mu$ as indicated by the above-mentioned curve.  For lower values of  $\mu$, the size of this region decreases until $\mu$ reaches the value for which horseshoe orbits begins to be stable. After these critical masses, the two domains shrink together but at a different rate.  The asymptotical estimates of the tadpole's width in this direction is of the order of $\mu^{1/2}$ (black solid line in Fig. \ref{fig:stabzp}), while an estimation for the horseshoe region is of the order of $\mu^{1/3}$ (black dashed line in Fig. \ref{fig:stabzp}, corresponding to the equation $ \Delta a = 0.47 \mu^{1/3}$)  has been fitted to the lower bound of the stable horseshoe region \citep[see][for more details]{RoPo2013}.
As a consequence, the stability domain of the horseshoe configurations becomes larger than the tadpole domain when the planetary masses tend to zero \citep{DeMu1981a}.
    
%
%
 

\section{One planet or two co-orbitals?}

In  some particular situations, co-orbital planets can be identified independently from the orbital libration: when both planets are transiting \citep{Ja2013} or when we combine data from transits with radial velocities \citep{FoGa2006}. However, in general the detection of co-orbitals requires  identifying the effect of the libration in the data. \citet{VoNe2014} showed that the TTV of only one of the  co-orbital planets is enough if the libration is large. \citet{LauCha2002} showed that the libration induced by co-orbital can have an important effect on the radial velocity of a star, while \citet{GiuBe2012} showed that co-orbitals can be mistaken for a single planet if the data span is short with respect to the libration period.

In the previous section we saw that co-orbital planets can be stable for large libration amplitudes, depending on the parameter $\mu$ (see Figures \ref{fig:orb} and \ref{fig:stabzr}). However, the libration period is always longer than the orbital period of the bodies (see the colour code in Figure \ref{fig:stabzr}). 
The faster $\zeta$ librates, the higher the chances of detecting the co-orbital bodies, because this reduces the time span needed to detect the libration. We write $P_\nu$ the period associated to the libration frequency $\nu$. The value of $P_\nu$ decreases when $\mu$ and $n$ increase (see equation (\ref{eq:nu_L}) and Figure \ref{fig:zr}). Therefore, high mass ratios and the proximity to the star maximize the detectability of co-orbitals, although high mass ratios also lead to the instability of most of the co-orbital configurations (Figure \ref{fig:stabzr}). Hereafter we consider that the time span of the observations is always longer than $P_\nu$.

%
%

\subsection{Signals induced by co-orbital planets}

\begin{figure}[h!]
\includegraphics[width=0.49\linewidth]{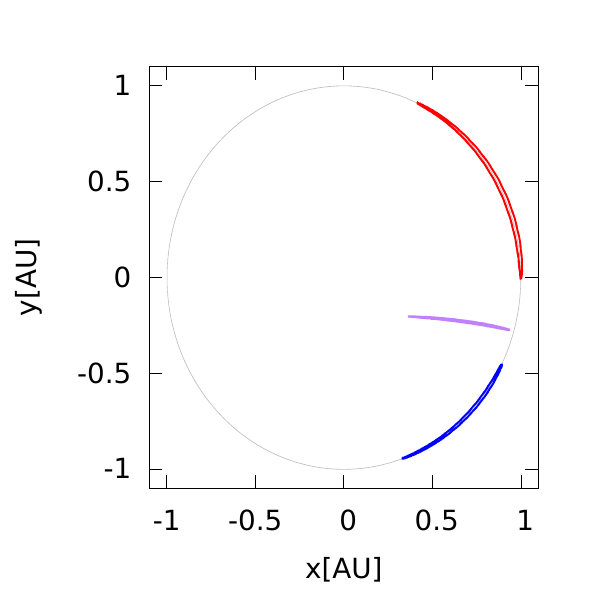}
\includegraphics[width=0.49\linewidth]{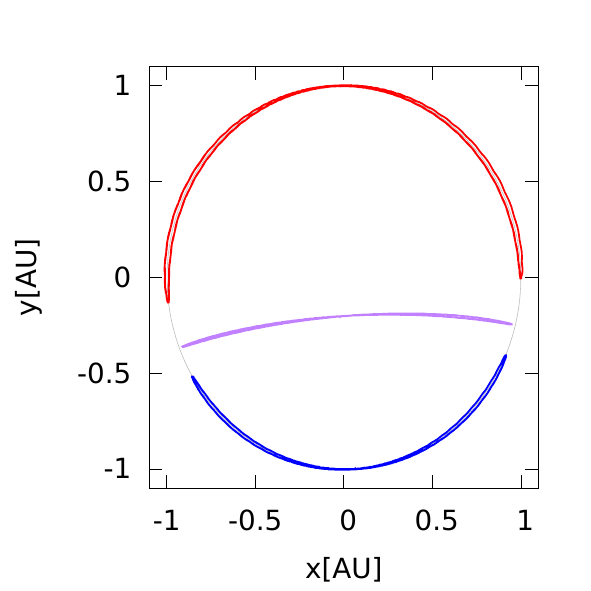}
\caption{Motion of the two co-orbital bodies (red and blue) and their barycentre (purple) in a co-rotating frame  with frequency $n$. Tadpole (left) and horseshoe (right). $\delta=0.6$. Here $\mu=2\,10^{-4}$ and the planets are located at $1\,AU$ from the star. By eliminating the influence of $n$, one can see the long-term motion of the barycentre of the planets. $P_\nu$ is the period of the periodic trajectories represented by the coloured lines. See the text for more details.}
\label{fig:schema_sym}
\end{figure}
%
%
%
%
\begin{figure}[h!]
\includegraphics[width=1\linewidth]{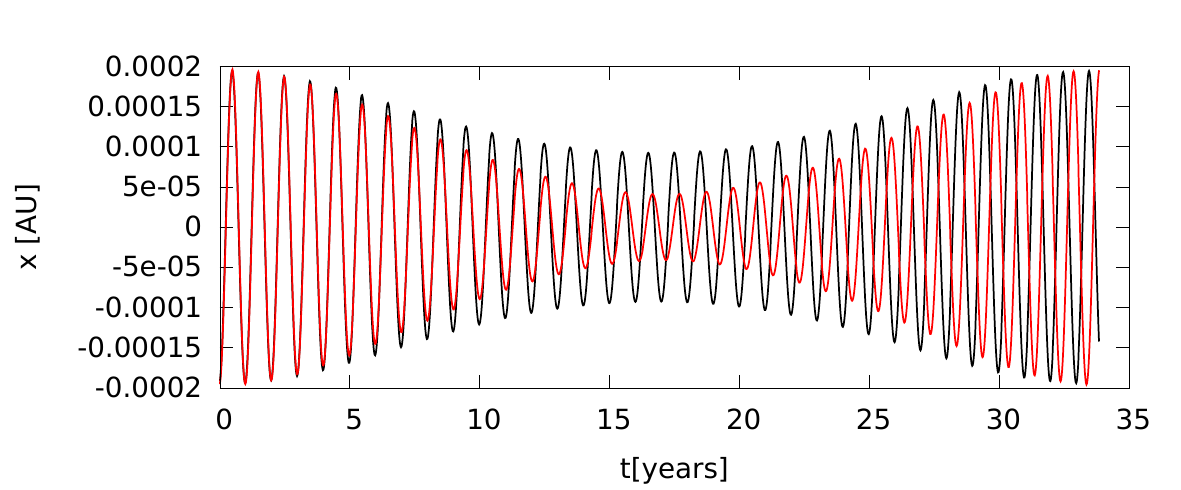}\\
\includegraphics[width=1\linewidth]{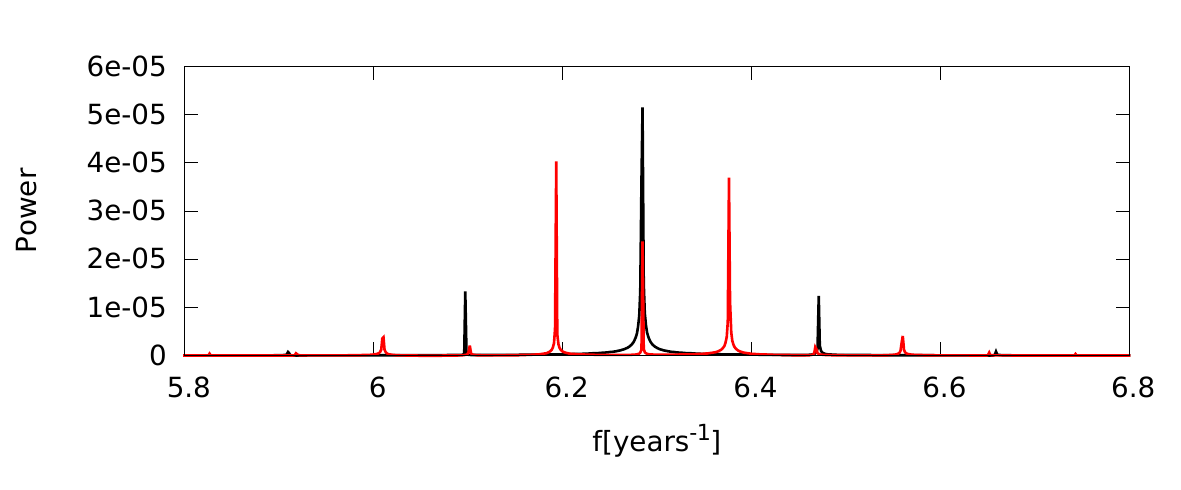}
\caption{Motion of the star in the configurations of Fig. \ref{fig:schema_sym} in the direction x in the inertial frame. In black is the tadpole orbit and in red the horseshoe. The top graph represents the evolution of the position of the star over time and the bottom  graph  its spectrum. In these examples, the libration period of the horseshoe orbits is about twice the period of the tadpole orbits. See the text for more details.}
\label{fig:VR}
\end{figure}

Most important observational techniques used to detect exoplanets (transits, radial-velocity, astrometry) are indirect, i.e. we do not directly observe the planets, but rather their effect on the host star.
In order to get an idea of the effect of the libration of co-orbital planets on the star, we take two examples of co-orbital configurations (see Figure \ref{fig:schema_sym}) with the following initial conditions: $\lambda_1=0^\circ$, $a_1=a_2=1\,AU$, $e_1=e_2=0$, $m_1=0.8\,10^{-4} M_\odot$ (red), and $m_2=1.2\,10^{-4} M_\odot$ (blue). In the left graph, $\zeta_0=25^\circ$, leading to a large amplitude tadpole orbit, and in the right graph, $\zeta_0=23^\circ$, leading to a horseshoe orbit. The position of the barycentre of the system composed of the two planets is represented in purple. With $\mu = 2 \times 10^{-4}$ and $\zeta_0$ near the separatrix, these two examples are at the limit of the stability domain, but give a clear idea of what we can expect.

In Figure \ref{fig:VR} we show the projection of the stellar orbit on the $x-$axis for these two configurations. We observe that the signal induced by the Keplerian  motion of the co-orbitals is indeed modulated over a period of libration of the resonant angle $\zeta$. This phenomenon was described by \citet{LauCha2002} in the case of a radial velocity signal. It is due to the oscillation, with a frequency $\nu$, of the distance between the barycentre of the two planets and the star, clearly visible in Figure \ref{fig:schema_sym}. The larger the amplitude of variation of $\zeta$, the larger the amplitude of modulation. For a given $\zeta_0$ value, the maximum oscillation amplitude is achieved when $\delta= 1/2$, that is, for $m_1=m_2$. In the horseshoe configuration, $\delta= 1/2$ leads the barycentre of $m_1$ and $m_2$ to pass by the position of $m_0$, periodically cancelling the signal.  

The bottom panel of Figure~\ref{fig:VR} shows the spectrum of those signals. The features of the spectrum of a modulated signal appear clearly: one peak located at the high frequency $n$ and harmonics located on both sides $n + p\nu$, where $p$ is an integer and $\nu$ is the frequency of the modulation. In general, the peaks located in $n$ and $n \pm \nu$ are the ones with the largest  amplitude. However, there is an exception when the signal is at the limit of the over-modulation, that is, when the peak located in $n$ disappears. This can happen only in the horseshoe configuration when $m_1 \approx m_2$. In this case, the main components of the spectrum would be two peaks separated by $2\nu$ and the system would then be easier to identify.
In this paper we  focus on the possibility of detecting the main three peaks.


\subsection{Motion of a star hosting co-orbital planets}

If the centre of mass of the system is at rest, the position of a star hosting two co-orbital planets is given by (in barycentric coordinates)
\be
\bold{r_0}= \mu [(1-\delta)\bold{r_1}+ \delta \bold{r_2}] \ ,
\label{eq:r0_v1}
\ee
where $\bold{r_1}$ and $\bold{r_2}$ are given by equations (\ref{eq:r_v1}).
Since $\zeta(t)$ is a periodic function with frequency $\nu$, we can expand the terms $e^{i \delta \zeta}$ in Fourier series as
\begin{equation}
\label{eq:deczet}
 e^{i\delta\zeta} = \sum_{p \in \mathbb{Z}}\ c_p(\delta,\zeta_0,t_0) e^{  i p \nu t} \ , 
\end{equation}
where $c_p(\delta,\zeta_0,t_0)$ is a complex coefficient. Replacing equation (\ref{eq:r_v1}) and (\ref{eq:deczet}) into equation (\ref{eq:r0_v1}), we get
\be
\bold{r_0}= \mu a \sum_{p \in \mathbb{Z}} |C_p| e^{i(p \nu t+ nt + \lambda_0 + \varphi_p)} \ ,
\label{eq:r0_v2}
\ee
with
\be
C_p=(1-\delta) \, c_p(\delta,\zeta_0,t_0)+ \delta \, c_p(\delta-1,\zeta_0,t_0) \ ,
\label{eq:Cp}
\ee
and
\be
\varphi_p = arg(C_p) \ .
\label{eq:Cpb}
\ee
For the velocity, we thus have (at order $0$ in $\nu$)
\be
\bold{\dot{r}_0}= i \mu a n \sum_{p \in \mathbb{Z}}| C_p| e^{i(p \nu t+ nt + \lambda_0 + \varphi_p)} \ .
\label{eq:rp0_v2}
\ee

For instance, if the observational data is acquired through astrometry, we get the projection of equation (\ref{eq:r0_v2}) on the plane of the sky, while for radial velocities we use the projection of equation (\ref{eq:rp0_v2}) in the line of sight.

The stellar motion can be expressed as the sum of a signal of frequency $n$, which we call the ``Keplerian component'', and other signals of frequency $n + p\nu$, which we call the ``modulating components''. For simplicity, we consider only the two main modulation components $p=\pm 1$, which are the ones with the largest amplitude, hence the ones that are easier to detect. We thus introduce the quantity $\vS(t)$, which represents a projection of $\bold{r_0}$ (equation (\ref{eq:r0_v1})) or $\bold{\dot{r}_0}$ (equation (\ref{eq:rp0_v2})) over an observable direction, restricted to its main two components,
\be
\vS (t) = \vK (t) + \vM (t) \ ,
\label{eq:st}
\ee
where
\be
\vK (t) = \bar{\vS}+\vS_0\,\cos(nt+\phi_0) \ ,
\label{eq:stk}
\ee
and
\begin{eqnarray}
\vM (t) &=& \vS_1\,\cos((n+\nu)t+\phi_1) \nonumber \\ && + \, \vS_{-1}\,\cos((n-\nu)t+\phi_{-1}) \ ,
\label{eq:stm}
\end{eqnarray}
where $\phi_0$, $\phi_1$, and $\phi_{-1}$ depend on $\varphi_{-1,0,1}$, $\lambda_0$, and the direction of the projection.
Our purpose is to check if the Keplerian signal that we have detected is modulated, and if our data can be approximated by a signal under the form $S(t)$. 

\subsection{Demodulation}
\label{sec:dem}

We assume that the Keplerian part of the signal is well determined ($\vS_0$ and $\bar{\vS}$ terms in equation (\ref{eq:stk})), otherwise it would be impossible to look for something else.
However, the modulating signal ($\vS_1$ and $\vS_{-1}$) can be hidden in the noise. In order to isolate the effect of the modulation, we suggest using a frequency mixing method similar to the one used in the demodulation of radio signals. This method is called ``superheterodyne'' and was introduced by \citet{Ar1914}. It consists in multiplying the modulated signal by a signal that has the same frequency as the carrier. As a result, we obtain a peak at the modulating signal's period in the spectrum.  
We propose using this method on data from co-orbital systems, but it can also be used on any other modulated signal produced by a different source \citep[e.g.][]{MoCo2008}.

We consider a set of $N$ observational data measurements.
We denote $t_k$ the time of each observation and $s_k$ the corresponding observed measurement.
First, we fit the data with a simple sinusoidal function that contains only the Keplerian part $\vK (t)$ (Eq.\,\ref{eq:stk}).
This provides us with an initial approximation for $\bar \vS$, $\vS_0$, $n$, and $\phi_0$. 
Then, we perform a transformation on the raw data $s_k$ to subtract the Keplerian part,
\be
s_k' = s_k - \vK(t_k) \ , 
\label{eq:demkp}
\ee
and then, to isolate the modulation frequency,
\be
\tilde{s}_k = s_k' \cos\,(nt_k+\phi) \ , 
\label{eq:dem}
\ee
where $\phi$ is an arbitrary phase angle. 
This modified data set can be fitted with a similarly modified function 
\begin{eqnarray}
\tilde \vS(t) & =& \left[\vS(t) - \vK(t) \right] \cos(n t+\phi) = \vM (t) \cos (n t + \phi) \nonumber \\
&=& \frac{\vS_1}{2} \cos((2n+\nu)t +\phi_1+\phi) \nonumber \\ 
&+& \frac{\vS_{-1}}{2} \cos((2n-\nu)t +\phi_{-1}+\phi) \nonumber \\
&+& \hat \vS_1 \cos (\nu t + \Delta \phi)
+ \Delta \vS \cos(\nu t + \phi - \phi_{-1})  \ ,  
\label{eq:demc}
\end{eqnarray}
where
\be 
\hat \vS_1 = \vS_1 \cos\,(\bar \phi - \phi) \ , \quad \Delta \vS = \frac{\vS_{-1} - \vS_1}{2} \ ,
\label{eq:demcxyz}
\ee
\be
\bar \phi =  \frac{\phi_{-1} +  \phi_1}{2} \ , \quad \Delta \phi = \frac{\phi_{-1} -  \phi_1}{2} \ . 
\ee
The libration frequency $\nu$ is now clearly separated from the Keplerian frequency $n$. As we will see in the following sections, we have $\Delta \vS \ll \vS_1$. The libration contribution can therefore be fitted by the term in $ \hat \vS_1 $, and the signal is maximized if we are able to choose $\phi = \bar \phi$. 
However, $\bar \phi$ is \textit{\emph{a priori}} unknown, so we propose  computing the $\tilde{s}_k$ for two values of $\phi$ dephased by $\pi/2$, for example $\phi=\phi_0$ and $\phi=\phi_0 + \pi/2$. By proceeding in this way, in the worst case we get $\bar \phi - \phi = \pi/4$, corresponding to a minimum amplitude of $\vS_1 / \sqrt{2} $.
Moreover, by taking the ratio of the fitted amplitudes with the two $\phi$ values, we can additionally estimate $\bar \phi $, and thus $\phi_{\pm 1} = \bar \phi \mp \Delta \phi$.

The initial determination of $n$ using equation (\ref{eq:stk})  always has an error $\epsilon_n$, which leads to the splitting of the libration term in $\nu$ into two terms in $\nu\pm \epsilon_n$. Since these two frequencies are very close to each other, the Fast Fourier Transform (FFT) usually shows a widened peak around $\nu$, preventing an optimal determination of $\nu$, $\vS_1$, and $\Delta \phi$. 
Therefore, once we have some estimations for these parameters, in the last step of the demodulation process, we return to the  original data set $s_k$, and directly fit it with the full equation $\vS(t)$ (equation (\ref{eq:st})), using the previously determined $\bar{\vS}$, $\vS_0$, $\vS_1$, $\vS_{-1}= \vS_1$, $n$, $\nu$, $\phi_0$, $\phi_1$, and $\phi_{-1}$ as initial values for the  fit.



\section{Detection using the radial-velocity technique}
\label{sec:rv}

In this section we apply the general method described previously to the case where the  data is acquired thorough the radial-velocity technique. In this case, the data corresponds to the projection of equation (\ref{eq:rp0_v2}) in the line of sight, given by an arbitrary direction $e^{i\theta}\sin\,I$ in the space \citep{MurCor2010}
\be
v_r(t)=\bold{\dot{r}_0} \cdot e^{i \theta} \sin\,I=  \alpha \sum_{p \in \mathbb{Z}} |C_p| \cos(p \nu t+ nt + \phi_p),
\label{eq:RV}
\ee
where 
\be
\alpha= \mu a n \sin I \ , \quad \mathrm{and} \quad
\phi_p=\varphi_p+ \pi/2-\theta +\lambda_0 \ .
\label{eq25}
\ee
We note that equation (\ref{eq:RV}) could also be the projection of equation (\ref{eq:r0_v2}) over a direction in the plane of the sky (for example in the case of an astrometric measurement). Within our approximation that would only change the value of the parameter $\alpha$. However, most of our results on the detectability do not depend on this parameter, thus hold true for any measurement technique. For reasons of clarity, we return  to the example of the radial velocity measurements.

Considering only the first harmonics of equation (\ref{eq:RV}), one can identify the RV signal to the equation (\ref{eq:st}), which is
\begin{eqnarray}
v_r(t) & =& \bar S + S_0\,\cos(nt + \phi_0) + S_{-1} \cos((n-\nu)t  + \phi_{-1}) \nonumber \\
&& \hskip.3truecm + S_1 \cos((n+\nu)t  + \phi_1) \ ,
\label{eq:RV_v2}
\end{eqnarray}
with $S_p= \alpha |C_p|$. 
We can therefore apply the demodulation process from section~\ref{sec:dem} to extract the orbital information from the observational data.
Our aim now is to determine which configurations can be detected for a given precision in the RV observations, and explain how to retrieve the orbital parameters from the $S_p$ and $\phi_p$ parameters.


\subsection{Detectability}
\label{sec:deti}

We introduce the following quantity
\be
 A_m=\frac{S_1+S_{-1}}{2 S_0} = \frac{|C_1|+|C_{-1}|}{2 |C_0|}\ ,
 \label{eq:Am}
\ee
which represents the power of the modulation terms with respect to the Keplerian term. 
When we search for co-orbital planets, the product $S_0 A_m$ must be distinguishable from the noise.

 \subsubsection{Detection near the Lagrangian equilibrium}
 \label{dnLe}

We consider a system in a tadpole configuration with a low libration amplitude. In this case we can use a linear approximation for $\zeta$ near the Lagrangian equilibrium. Within this approximation, we can obtain an explicit expression for $v_r(t)$ in terms of the orbital parameters. We  introduce the small parameter $z=\zeta_0 - \pi/3$ and write
\be
\zeta(t) =  \frac{\pi}{3} + z \cos\,(\nu(t-t_0)).
\label{eq:z_v1}
\ee

At first order in $z$ and using equation (\ref{eq:rp_v1}), the derivative of equation (\ref{eq:r0_v1}) becomes %
\be
\begin{split}
\bold{\dot{r}_0}= - i \mu a n  \Big[(1-\delta) \big(1+i \delta z \cos\,(\nu(t-t_0)) \big) \\
 + \delta  \big(1+ i (1-\delta) z \cos\,(\nu(t-t_0))\big) e^{-i \frac{\pi}{3}} \Big] e^{i(nt+ \lambda_0+\delta \frac{\pi}{3})}.
\end{split}
\label{eq:rp0_v1}
\ee
Following equation (\ref{eq:RV}), we project  equation (\ref{eq:rp0_v1}) in the line of sight, and identify the terms appearing in equation (\ref{eq:RV_v2}) as
\be
S_0= \alpha \sqrt{1-\delta(1-\delta)} \ ,
\label{eq:s0l}
\ee
\be
S_1=S_{-1}= \alpha \frac{ \sqrt{3}\delta (1-\delta)}{2} \, z \ ,
\label{eq:s1l}
\ee
which allow us to compute $A_m$ as well:
\be
A_m=\frac{\sqrt{3}}{2}\frac{\delta(1-\delta)}{\sqrt{1-\delta(1-\delta)}} \, z \ .
\label{eq:Aml}
\ee

When $m_0 \gg m_2 \ge m_1$ the modulation terms can be simplified as 
\be
S_1 =S_{-1} \approx \frac{\sqrt{3}}{2}\frac{\beta}{m_0} z a n \sin I \ ,
\ee
where $\beta = m_1 m_2 / (m_1 + m_2)$ is the reduced mass of the planets' subsystem.
We thus see that the power in these terms is proportional to $\beta$ and to the angular separation from the Lagrangian equilibrium $z$.
The detection is therefore maximized for large libration amplitudes and planets with large similar masses ($m_1 \approx m_2$).
Nevertheless, we note that for planetary systems with mass ratios very different from one, for instance, $m_1 / m_2 \ll 1$, the reduced mass converges to the mass of the smaller planet ($\beta = m_1$ in this case), while for equal mass planets it converges to $\beta = m_1 / 2$.
As a consequence, although planets with large equal masses are easier to detect than planets with small equal masses, a small mass planet is two times easier to detect if it is accompanied by a large mass planet rather than another small similar mass planet.

 \subsubsection{Detection in any tadpole or horseshoe configuration}
 \label{sec:detg}
 
For large libration amplitudes, we cannot have an explicit expression for $S_p$ with respect to the orbital parameters. Nevertheless, similarly to the linear case, we can prove that $A_m$ and $|C_0|$ depend only on $\zeta_0$ and $\delta$. Indeed, since $c_p$ are Fourier coefficients of the expression of $e^{i\delta\zeta}$, see equation (\ref{eq:deczet}), we can write
%
\be
c_p(\delta,\zeta_0,t_0)=\frac{\nu}{2\pi}\int^{\pi/\nu}_{-\pi/\nu}e^{i\delta\zeta(t-t_0)} e^{-ip\nu t} dt,
\label{eq:cp_v1}
\ee 
or, in terms of $\tau$ (see section \ref{sec:coo}),
\be
c_p(\delta,\zeta_0,t_0)=\frac{\tilde{\nu}}{2\pi}\int^{\pi/\tilde{\nu}}_{-\pi/\tilde{\nu}}e^{i\delta\tilde{\zeta}(\tau-\tau_0)} e^{-ip\tilde{\nu} \tau} d\tau,
\label{eq:cp_v1}
\ee 
where $\tau_0=t_0/(n\sqrt{\mu})$. Since $\tilde{\nu}$ depends only on $\zeta_0$, it turns out that
\be
c_p(\delta,\zeta_0,t_0)= c_p(\delta,\zeta_0,0) e^{-ip\tilde{\nu} \tau_0}= c_p(\delta,\zeta_0,0) e^{-ip\nu t_0}.
\label{eq:cp_v5}
\ee 
As a result, the dependence of  $c_p$ on $\tau_0$ is explicit. Using the definition of $C_p$ given by equation (\ref{eq:Cp}), we see that $|C_p|$, and consequently $A_m$, do not depend on $t_0$.

The dependence of $A_m$ and $|C_0|$ on the parameters ($\delta$, $\zeta_0$) is shown in Figures \ref{fig:modphaT} and \ref{fig:A0T} for tadpole configurations and in Figure \ref{fig:A0modHS} for horseshoe configurations. These figures were obtained by integrating the differential equation (\ref{eq:sol_orb_zeta}) satisfied by $\zeta$, with initial conditions $(\zeta(0), \dot{\zeta}(0))=(\zeta_0, 0)$. The outputs of these integrations were then replaced into the expression of $\bold{\dot{r}_0}$ for a given set of $\delta$ (Eq.\,\ref{eq:rp_v1}). For each simulation, the spectrum of a projection of $\bold{\dot{r}_0}$ has been computed in order to get the value of the displayed quantities. These quantities have also been computed from three-body direct integrations, which give the same results.

The RV signal that we obtain for the general cases follows the trends of the linear approach.
For  given values of  $a$, $\mu$, and $\delta$, the detectability of a co-orbital system increases as the amplitude of the libration of $\zeta$ increases, i.e. when $\zeta_0$ decreases. This is still true when $\zeta_0$ crosses the separatrix. 
When $\delta$ tends to $1$ or $0$, the modulation peak disappears and the signal is similar to the one induced by a single planet. For a given $\zeta_0$, $A_m$ reaches its maximum when $\delta = 1/2$. In the horseshoe case, the modulating terms have higher amplitudes than the Keplerian term for $0.35 \lesssim \delta \lesssim 0.65$, the Keplerian term being cancelled when $\delta$ tends to $1/2$.

We showed at the end of the previous section that a planet of mass $m_1$ (fixed) will be easier to detect if its co-orbital companion is significantly more massive ($m_2 \gg m_1$), rather than $m_2 \approx m_1$. This holds true in the horseshoe domain,  as shown in Appendix \ref{ap:am}.

\subsubsection{Detectability for a given data set}
 
While searching for co-orbital companions of an already detected planet, it is possible to put some constraints on what we can expect to observe, based on the observational limitations.
In addition to the main Keplerian signal, characterized by $K_0$ and $P_n$, we also know the time span of the observations, $T$, and the precision of the instrument, $\epsilon$. 

The modulation signal of a co-orbital configuration is detectable if $A_m K_0 > \epsilon$ (Eq.\,\ref{eq:Am}). Thus, the detection of a co-orbital companion can only occur for 
\be
\frac{1}{A_m} < \frac{K_0}{\epsilon} \ . \label{dAm}
\ee
We also know that the libration period $P_\nu$ is proportional to the orbital period $P_n$ (Eq.\,\ref{eq:nut}, Fig.\,\ref{fig:zr}). 
One complete libration period can only be contained in the data when $P_\nu > T$, therefore
\be
\frac{P_\nu}{P_n} > \frac{T}{P_n} \ . \label{dpnu}
\ee
The parameter $A_m$ depends on $\delta$ and $\zeta_0$, while the ratio $P_\nu/P_n$ depends on $\zeta_0$ and $\mu$.
The detectability of a co-orbital configuration therefore depends on the mass of both planets and on the libration amplitude.

\begin{figure}[ht]
\includegraphics[width=1\linewidth]{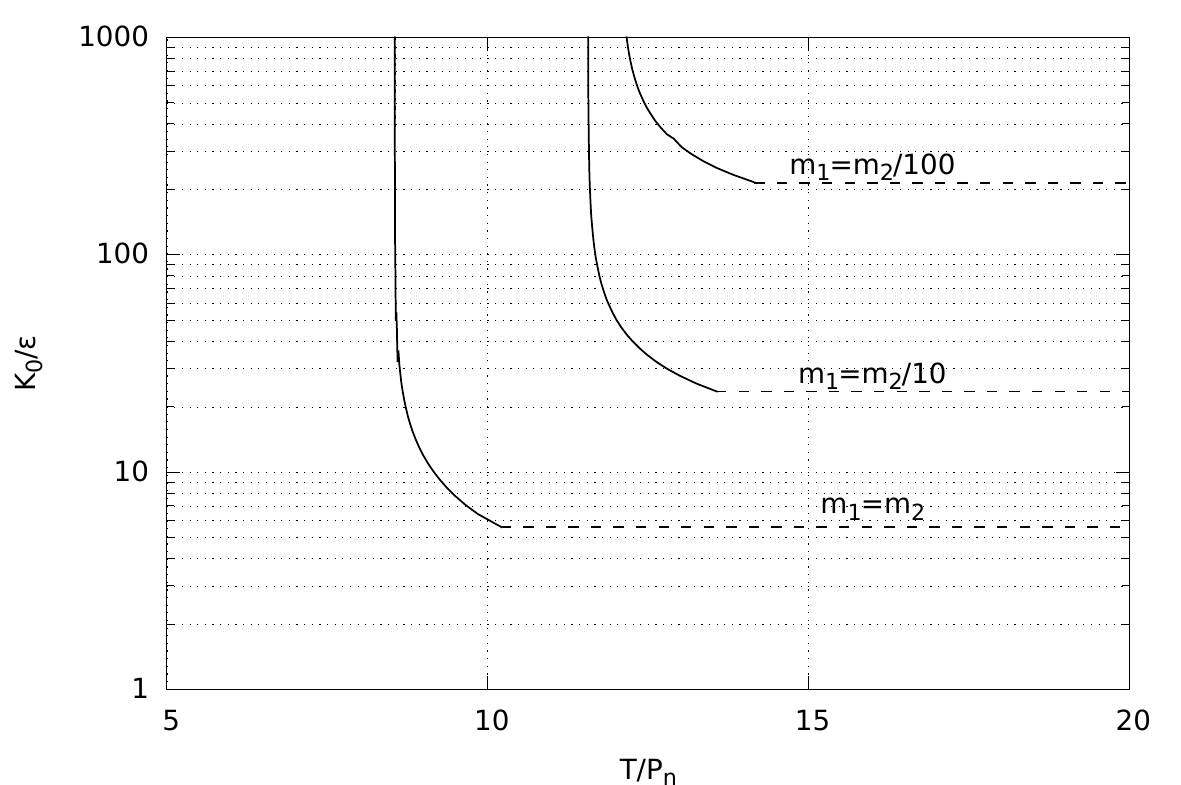}
\caption{Detectability of a co-orbital companion for $m_2/m_0 \approx 10^{-3}$. For a  given data set ($K_0/\epsilon$,$T/P_n$), co-orbital companions with a mass $m_1$ can only be detected if they lie above the respective threshold limit.}
\label{fig:Detect}
\end{figure}

In Figure~\ref{fig:Detect} we show the ratio $K_0/\epsilon$ as a function of the ratio $T/P_\nu$, which correspond to the observable quantities.
We denote $m_2$ the most massive of the two planets (which is the main contributor to $K_0$ and $\mu$), and $m_1$ the mass of the less massive planet that we are looking for.
We fix $m_2/m_0=10^{-3}$ (which is near the maximum value allowed for the stability of co-orbital systems) and show the detection limits for three different values of $m_1$.
Co-orbital companions below each threshold limit can be ruled out.

These detection limits are constrained by the observational limitations ($\epsilon$ and $T$), but also by the stability of the co-orbital systems, which is parametrized by the values of $\zeta_0$.
As $\zeta_0 \rightarrow \pi/3$ (Lagrange point, with no libration amplitude) or $m_1/m_2 \rightarrow 0$, we have that $K_0/\epsilon > 1/A_m \rightarrow \infty$.
On the other hand, as $\zeta_0 \rightarrow 0$, the chances of detection increase (the libration amplitude increases), but the system also tends to become unstable (Fig.\,\ref{fig:stabzr}).

\subsection{Characterization of the co-orbital system} 
\label{sec:inv}
The orbits of the co-orbital planets are fully characterized by the quantities $n$, $\nu$, $a$, $\zeta_0$, $\lambda_0$, $t_0$, and $\sin I$. In addition, assuming that the mass of the star is known, we can determine the mass of the planets through $\mu$ and $\delta$. The frequencies $n$ and $\nu$ are directly obtained when we fit the data with our model (Eq.\,\ref{eq:RV_v2}), while $a$ is obtained by the third Kepler law from $n$. Since $\tilde{\nu}$ depends only on $\zeta_0$ (Fig.\,\ref{fig:zr}), for each configuration there is a bijective map that links $\mu$ and $\zeta_0$ given by
\be
\sqrt{\mu}= \frac{\nu}{\tilde{\nu}(\zeta_0)n} \ ,
\label{eq:smu}
\ee
where $\tilde{\nu}$ is defined by equation\,(\ref{eq:nut}). We are thus left with five parameters, $\zeta_0$ (or $\mu$), $\delta$, $\lambda_0$, $t_0$, and $\sin I$, that need to be determined in order to characterize the system.

We can start looking for the shape of the orbit rather than the exact trajectories of the planets as a function of time. Therefore, we ignore by now all quantities that depend on $\lambda_0$ and $t_0$, i.e. we restrict our analysis to $A_m$ (Eq.\,\ref{eq:Am}) and $|C_0|$ (Eq.\,\ref{eq:cp_v5}).

We define the quantity $\Psi$ as
\be
 \Psi= 2(\bar \phi - \phi_0) = \phi_1+ \phi_{-1} - 2\phi_0 \ ,
\label{eq:psi0}
\ee 
with $\phi_p=\lambda_0+ arg(C_p)+\pi/2-\theta$ (Eq.\,\ref{eq25}). Thus
\be
\begin{split}
 \Psi=  arg(C_1(\delta,\zeta_0,t_0))+  arg(C_{-1}(\delta,\zeta_0,t_0))\\
  - 2\,arg(C_0(\delta,\zeta_0,t_0)) \ .
\end{split}
\label{eq:psi}
\ee 
From equation (\ref{eq:cp_v5}), we know that $arg(C_p(\delta,\zeta_0,t_0))=arg(C_p(\delta,\zeta_0,0))-p\nu t_0$. Hence $\Psi$ depends only on $\zeta_0$ and $\delta$. One can show that any quantity defined as a function of $\phi_p$ with $p \in \{-1,0,1\}$ and independent of $t_0$ and $\lambda_0$ is a function of $\Psi$. \\

The parameters $A_m$, $|C_0|$, and $\Psi$ evolve in a different way depending on the orbital configuration of the system (tadpole or horseshoe).
We thus need to split our analysis for these two different configurations.

\subsubsection{Characterization near the Lagrangian equilibrium}

In  the linear case, we can entirely determine the trajectories of the co-orbitals analytically. According to equations (\ref{eq:s0l}) and (\ref{eq:s1l}), the amplitudes $S_0$ and $S_1=S_{-1}$ depend on $\alpha$, $\zeta_0$, and $\delta$. By identifying the phases angles appearing in equation (\ref{eq:RV_v2}) to the data and then comparing with expression (\ref{eq:rp0_v1}), we get three additional equations
\be
\phi_0=  \lambda_0+\delta \frac{\pi}{3}- \arctan(\frac{\sqrt{3}\delta}{2-\delta}) \ ,
\label{eq:phi0l}
\ee
 and 
\be
\phi_{\pm 1}=  \lambda_0+\delta \frac{\pi}{3} -\frac{\pi}{6} \mp \nu t_0 \ . 
\label{eq:phil}
\ee

These three equations, combined with the equations (\ref{eq:s0l}) and (\ref{eq:s1l}) lead to a system of five equations of the form $(S_0,S_1,\phi_0,\phi_1,\phi_{-1})=F(\alpha, \delta, \zeta_0, \lambda_0,t_0)$, where $F$ is a non-linear function of the five unknown parameters. We can thus get an explicit expression for these parameters from the observational data. Then, the expression of $\nu$ near the Lagrangian equilibrium (Eq.\,\ref{eq:nu_L}) can be used to get the value of $\mu$. Finally, the inclination $I$ can be deduced from the definition of $\alpha$ (Eq.\,\ref{eq25}):
\be
\sin I= \frac{\alpha}{\mu a n} \ .
\label{eq:sini}
\ee
We can thus remove the classic $\mu \sin I$ degeneracy in this case and fully determine the exact masses of the planets and their trajectories in space.

Replacing expressions (\ref{eq:phi0l}) and (\ref{eq:phil}) for $\phi_p$ in the expression of $\Psi$ (Eq.\,\ref{eq:psi0}) gives
\be
 \Psi = 2\arctan(\frac{\sqrt{3}\delta}{2-\delta}) - \frac{\pi}{3} \ ,
\ee
i.e. near the Lagrangian equilibrium $\Psi$ only depends on $\delta$.
Since $ 0 \le \delta \le 1 $, we have $ \Psi \in [-\pi/3, \pi/3] $, and for $\delta = 1/2$ we get $\Psi = 0$, which corresponds to equal mass planets.

\begin{figure}[ht]
\includegraphics[width=1\linewidth]{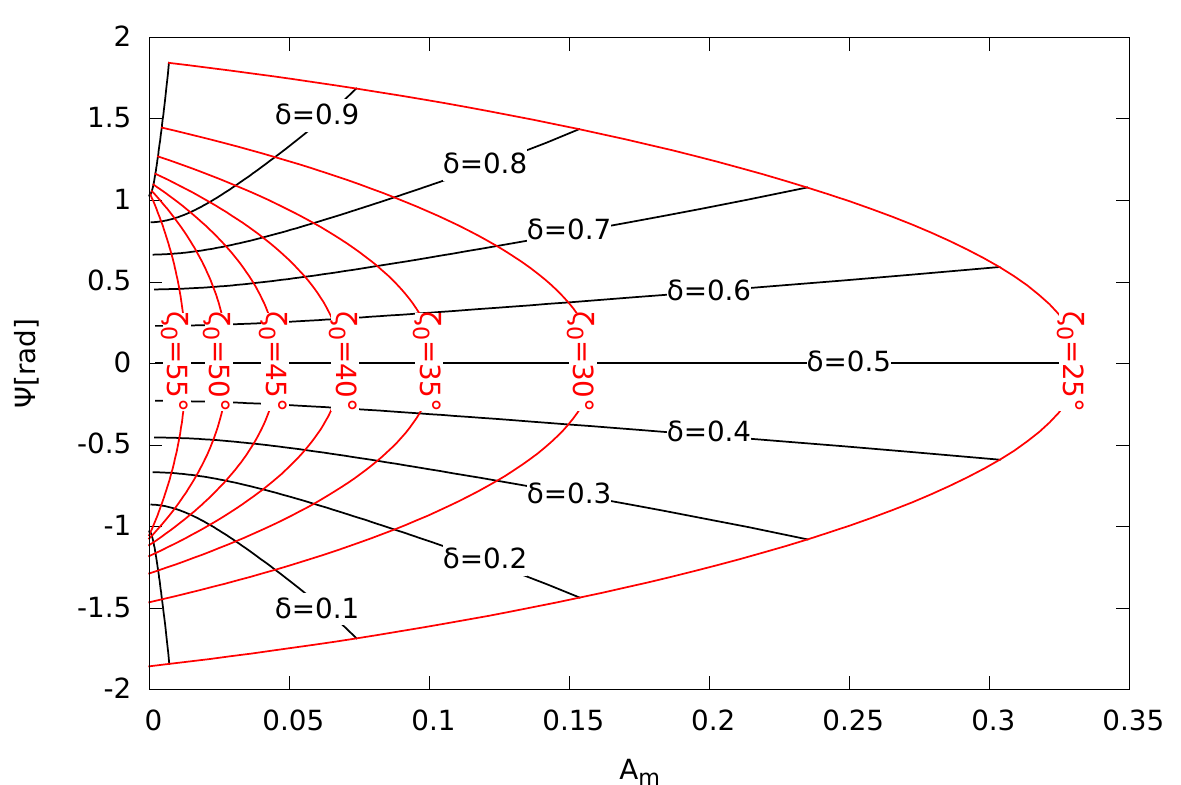}
\caption{Level curves of $\delta$ (black) and $\zeta_0$ (red) for the tadpole configuration, with respect to $A_m$ and $\Psi$. See the text for more details.}
\label{fig:modphaT}
\end{figure}

\begin{figure}[ht]
 \includegraphics[width=1\linewidth]{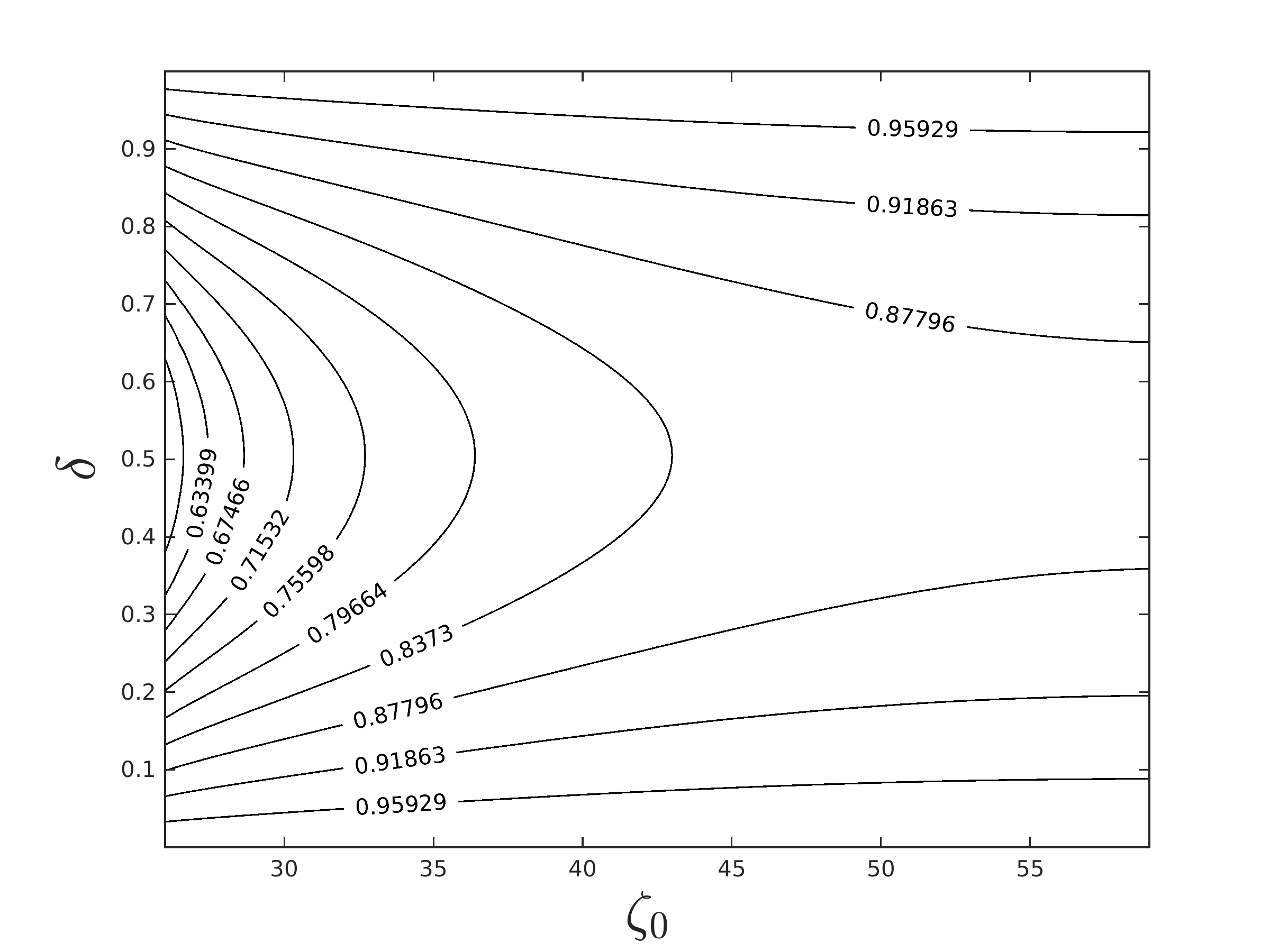}
\caption{Level curves of $|C_0|$ for the tadpole configuration with respect to $\zeta_0$ and $\delta$. See the text for more details.}
\label{fig:A0T}
\end{figure}
 
 \subsubsection{Large amplitude tadpole orbits}
\label{sss:inv}

As discussed in section~\ref{sec:detg}, for large libration amplitudes it is not possible to obtain an explicit expression for the orbital parameters from the $S_p$ terms.
The same applies to the  phase angles $\phi_p$. 
However, for tadpole configurations it is still possible to inverse the problem using implicit functions and to fully characterize the orbits from the modulation terms in equation (\ref{eq:RV_v2}). 


In Figure \ref{fig:modphaT}, we show iso-values of the parameters $\zeta_0$ and $\delta$ with respect to the quantities $A_m$ and $\Psi$ (see section \ref{sec:detg} for more details). For tadpole orbits, we see that  each couple ($A_m$, $\Psi$) corresponds to a unique couple ($\zeta_0$, $\delta$). One can thus determine the values of $\zeta_0$ and $\delta$ directly from $A_m$ and $\Psi$.

We also know that $|C_0(\delta,\zeta_0)|$ depends only on $\zeta_0$ and $\delta$ (see section \ref{sec:detg}). In Figure \ref{fig:A0T} we show iso-values of $|C_0|$. Since $S_0=\alpha |C_0|$, we can directly obtain the value of $\alpha$ from ($\zeta_0$, $\delta$), and hence from ($A_m$, $\Psi$).
We can thus determine $\sin I$ (Eq.\,\ref{eq:sini}), since $\mu$ is linked to $\zeta_0$ through expression (\ref{eq:smu}). The parameters $\delta$, $\zeta_0$, $\mu$, and $\sin I$ are then fully determined for the tadpole configuration.

Finally, similarly to the linear case (section~\ref{dnLe}), for a given $\delta$ one can show that $\phi_0$ is a bijective map for $\lambda_0 \in [0,2\pi/n[$, and $\phi_1-\lambda_0$ is a bijective map for $t_0 \in [0,2\pi/\nu[$ (see equations (\ref{eq:phi0l}) and (\ref{eq:phil})). The values of $\lambda_0$ and $t_0$ are therefore determined by the values of $\phi_0$ and $\phi_1$. Then, one can use  equations (\ref{eq:lamb}) and (\ref{eq:sol_orb_zeta}) to obtain the orbital parameters of the co-orbitals.

 \subsubsection{Horseshoe orbits}
\label{sss:invhs}

For the horseshoe configuration, it is also not possible to obtain explicit expressions for the orbital parameters from the $S_p$ terms. However, a symmetry in $\zeta$ allows us to compute this (see Appendix \ref{ap:psi}):
 \be
\Psi= arg(C_1)+arg (C_{-1}) - 2arg(C_0)  = \pi \ .
\label{eq:PmHS}
\ee
Since $\Psi$ is constant in horseshoe configurations, we cannot use it to get an additional constraint on the couple $(\delta, \zeta_0)$.

\begin{figure}[h!]
\includegraphics[width=0.49\linewidth]{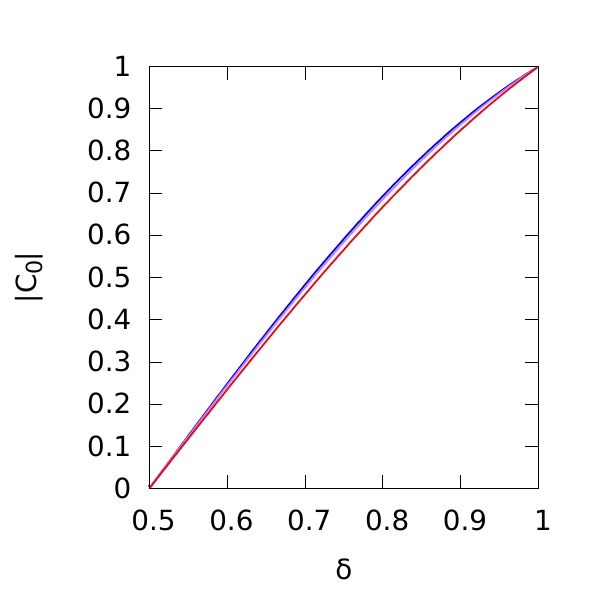}\includegraphics[width=.49\linewidth]{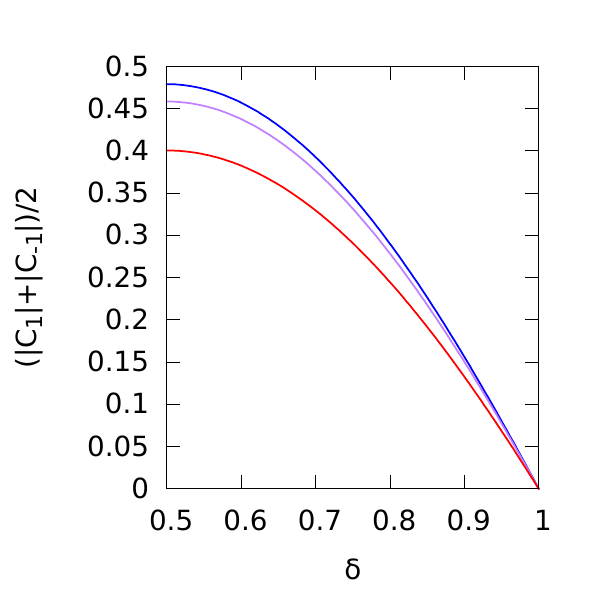}
\caption{\textit{Left:} $|C_0|$  with respect to $\delta$ in the horseshoe configuration. As $A_m(\delta=1/2)=+\infty$, we plot the quantity $A_m|C_0|$. \textit{Right:} $A_m|C_0|$ versus $\delta$ in the horseshoe configuration. These quantities are symmetric with respect to $\delta=0.5$. red: $\zeta_0=23^\circ$, purple: $\zeta_0=19^\circ$, blue: $\zeta_0=15^\circ$. See the text for more details.}
\label{fig:A0modHS}
\end{figure}
In Figure~\ref{fig:A0modHS} we plot $|C_0|$ and $A_m|C_0|=(|C_1|+|C_{-1}|)/2$ versus $\delta$ (see section~\ref{sec:detg} for more details). The graphs are symmetric in $\delta=1/2$. One can see that these quantities vary significantly with $\delta$, but are are almost constant in regard to $\zeta_0$ (different colour curves in Fig. \ref{fig:A0modHS}), except near $\delta=1/2$ for $A_m|C_0|$. 
Thus, 
we can assume an average value for $\zeta_0$ in the horseshoe domain. From this average value, we get approximated values of $\delta$ and $\alpha$ by knowing $A_m$ and $|C_0|$. Then, we can obtain approximated values for the parameters $t_0$ and $\lambda_0$ from $\phi_0$ and $\phi_1$, as explained in the tadpole case. However, the degeneracy in $\mu \sin I $ remains, because of the strong dependence of $\tilde{\nu}$ on $\zeta_0$ in the horseshoe domain (see Fig. \ref{fig:zr}).
One of the ways to get this information is to consider higher order harmonics in the expansion of the radial velocity, equation (\ref{eq:RV}). However, as these harmonics are about $10$ times smaller than $S_1$, much more accurate data is required.

\subsubsection{Tadpole or horseshoe?}
\label{sec:det}

Since the method that we use to determine the orbital parameters of a co-orbital system depends on its configuration (tadpole or horseshoe), it is legitimate to ask whether it is  possible to know the configuration type before we choose one method or another for reducing the observational data.\\

Once the signature of a co-orbital system is detected (by the observation of a modulation in the radial-velocity data) we can compute $\Psi$ from equation (\ref{eq:psi}). One can see from equation (\ref{eq:PmHS}) that $\Psi=\pi$ in the horseshoe configuration, while $\Psi \in [-2,2]$ in the tadpole configuration (Fig. \ref{fig:modphaT}). Since the domains for $\Psi$ are exclusive in the different configurations, by computing $\Psi$ we can immediately distinguish between a horseshoe and a tadpole configuration.\\

When the detected signal is at the limit of the instrumental precision, the phases can be improperly determined. In this case, one can always compute $A_m$ using expression (\ref{eq:Am}). As shown in Appendix \ref{ap:am}, $A_m$ ranges within $[0,+\infty[$ in the horseshoe configuration. In the tadpole configuration, $A_m$ reaches its maximum value for $\delta=1/2$ and $\zeta_0$ near the separatrix. We can see in Fig. \ref{fig:modphaT} that this quantity remains below $1/3$. Therefore, $A_m > 1/3$ is also a sufficient condition to know that a co-orbital system is in a horseshoe configuration.


\subsection{Application to synthetic data}

We now apply the methods developed in the previous sections to two concrete situations of stars hosting a pair of coobital planets in quasi-circular orbits, one for tadpole and another for horseshoe orbits.
In Table \ref{tab:pari} we list the initial osculating orbital elements for these two hypothetical systems orbiting a solar-mass star.
We then generate synthetic radial-velocity data for these systems by numerically integrating the equations of motion using an n-body model. In order to create a realistic data set, we use the same observational dates taken for the HD\,10180 system \citep{LoSe2011} to simulate the acquisition days, and associate with each measurement a Gaussian error with $\sigma=1$~m/s.
These synthetic data sets contain 160 measurements spanning 4600~days and correspond to an instrumental precision of $\sim 1$~m/s.
The orbital periods of the planets are around $11.5$~days in both examples, such that we can observe at least three complete libration cycles over the length of the observations.


\begin{table}[h]
\begin{center}
\caption{Osculating orbital elements for a given date of two hypothetical co-orbital systems orbiting a solar-mass star.}
\label{tab:pari}
\begin{footnotesize}
\begin{tabular}{|c|cc|cc|}
\hline
 &\multicolumn{2}{c|} {tadpole} &\multicolumn{2}{c|} {horseshoe}  \\ 
param. &  planet 1 & planet 2  &  planet 1 & planet 2 \\
\hline
\hline
$m$\,[$M_{\oplus}$]\dotfill & $200$ & $100$ & $17.15$ & $3.00$ \\
$a$\,[au]\dotfill & 0.0987 & 0.1013 & 0.1 & 0.1 \\
$\lambda$\,[deg]\dotfill & 0 & 300 & 0 & 339 \\ 
$e$\,\dotfill & 0.05 &0.05 & 0 & 0 \\ 
$\varpi$[deg]\dotfill & 0 & 300 & 0 & 0 \\ 
$I$\,[deg]\dotfill & 60 & 60 & 90   & 90 \\ 
$\bar{S}$\,[$km.s^{-1}$]\dotfill &\multicolumn{2}{c|}  {$6.500$} & \multicolumn{2}{c|} {$6.500$}  \\
\hline
$\bar{a}$\,[au]\dotfill &\multicolumn{2}{c|}  {$0.09955$} & \multicolumn{2}{c|} {$0.10000$}  \\
$\zeta_0$\,[deg]\dotfill &\multicolumn{2}{c|}  {$37.00$} & \multicolumn{2}{c|} {$21.00$}  \\
$\delta$\,\dotfill &\multicolumn{2}{c|}  {$0.3333$} & \multicolumn{2}{c|} {$0.1488$}  \\
\hline
\end{tabular}
\end{footnotesize}
\end{center}
\end{table}

\subsubsection{Tadpole orbits}

\begin{figure}[h!]
\begin{center}
\vspace{.5cm}
\hspace{5cm} (a) \hspace{7cm} (b) \vspace{7cm}\\
\hspace{5cm} (c) \hspace{7cm} (d) \vspace{-8.7cm}

 \includegraphics[width=.49\linewidth]{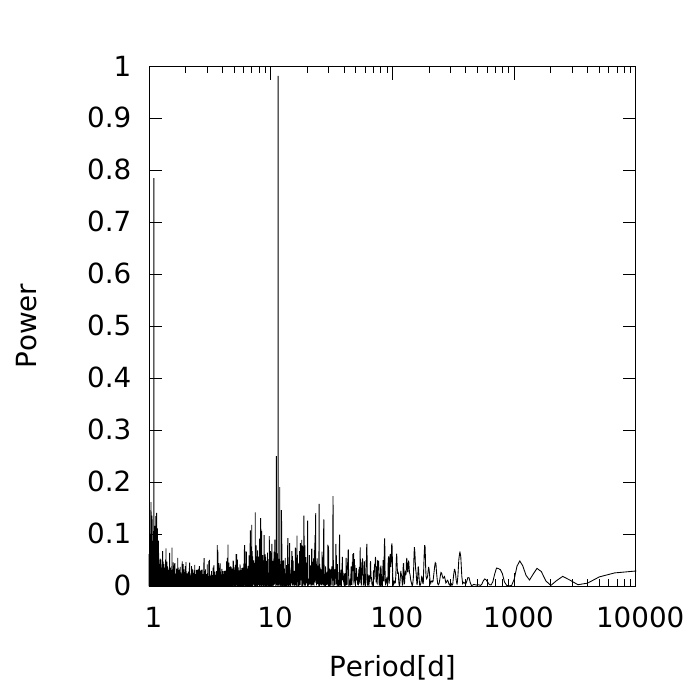}
 \includegraphics[width=.49\linewidth]{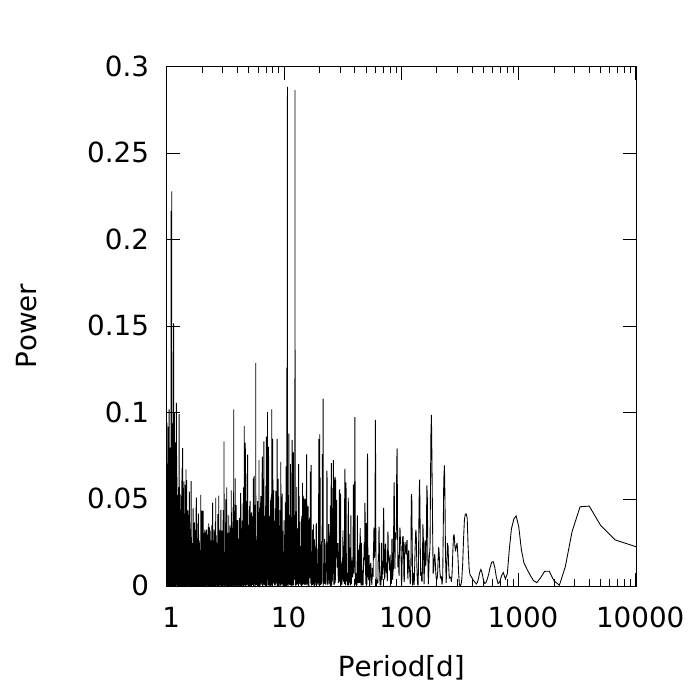} \\
\includegraphics[width=.49\linewidth]{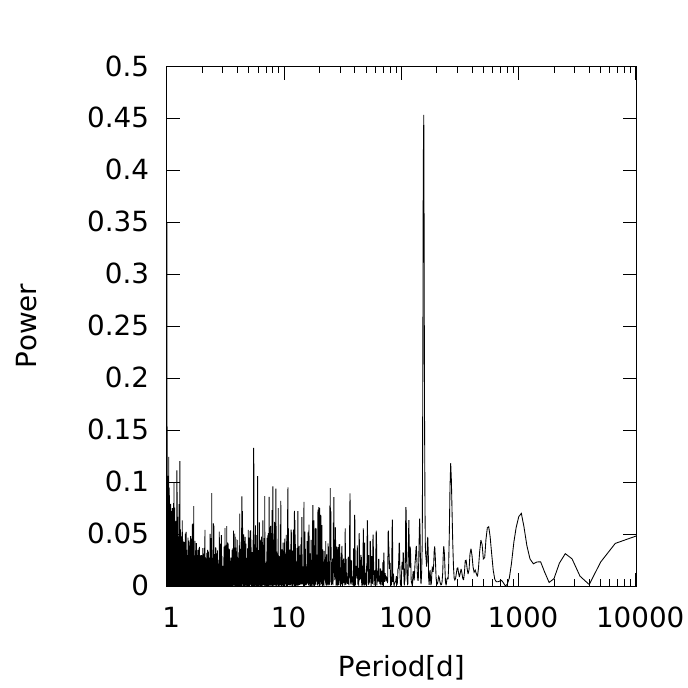}
\includegraphics[width=.49\linewidth]{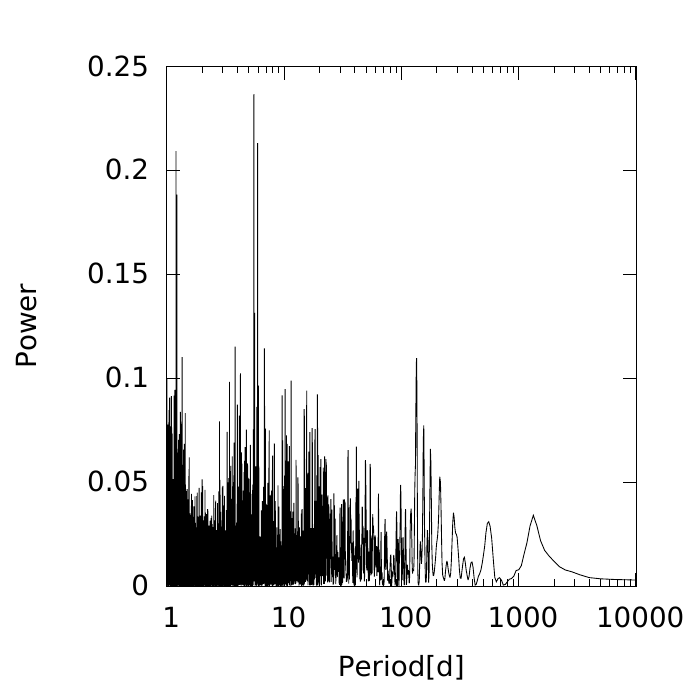}
\caption{Periodograms of the synthetic radial velocity of the tadpole configuration presented in Table \ref{tab:pari}. \textit{(a)} raw data $s_k$; \textit{(b)} modified data $s_k'$, after the subtraction of the Keplerian signal (Eq.\,\ref{eq:demkp}); \textit{(c)} modified data $\tilde s_k$ with $\phi = \phi_0$ (Eq.\,\ref{eq:dem}); and \textit{(d)} modified data $\tilde s_k$ with $\phi = \phi_0 + \pi/2$ (Eq.\,\ref{eq:dem}).}
\label{fig:fft_t}
\end{center} 
\end{figure}

Our  tadpole system is composed of two Saturn-like planets at $0.1$~au (comparable masses and eccentricities). The individual RV amplitudes of both planets are $K \sim 10$~m/s, well above the instrument precision.
Therefore, the signatures of the planets can be easily identified in the data, and we use this example to illustrate how to retrieve the complete set of orbital parameters listed in Table~\ref{tab:pari} with our method.

In Figure~\ref{fig:fft_t}\,(a), we show a generalized Lomb-Scargle periodogram of the radial velocity data \citep{ZeKu2009}. The Keplerian component of the signal with a period $P_n\approx 11 $~days can clearly be identified. We fit the raw data with a Keplerian function (Eq.\,\ref{eq:stk}) and obtain an initial estimation for $P_n \approx 11.46$~days, $\bar S \approx 6.5$~m/s, $S_0 \approx 61.1$~m/s, and $\phi_0 \approx 341.6^\circ$.
We then subtract the Keplerian contribution to the data and obtain a modified data set $s_k'$ (Eq.\,\ref{eq:demkp}).
In Figure~\ref{fig:fft_t}\,(b), we show a periodogram of this modified data.
We observe that the main peak with a period of approximately $11$ days is replaced by two nearby smaller peaks.
This is a clear indication of the presence of a modulation, each peak corresponding to the $n\pm\nu$ terms (Eq.\,\ref{eq:RV_v2}).

In order to better determine the libration frequency, we modify  the data again following expression (\ref{eq:dem}) adopting $\phi = \phi_0 = 341.6^\circ$ and $\phi = \phi_0 + \pi/2 = 71.6^\circ$.
In Figures~\ref{fig:fft_t}\,(c) and \ref{fig:fft_t}\,(d) we show the periodograms of $\tilde s_k$ for these two transformations, respectively.
In both transformations we observe that the peak around $11$ days is replaced by some power at the periods near $5$ and $150$ days, corresponding to the frequencies $2n$ and $\nu$, respectively (Eq.\,\ref{eq:demc}).
However, while for $\phi = \phi_0$ the maximum power is observed for $\nu$ (Fig.\,\ref{fig:fft_t}\,c), for $\phi = \phi_0 + \pi/2$ it is observed for $2n$ (Fig.\,\ref{fig:fft_t}\,d).
From expression (\ref{eq:demcxyz}), we see that the amplitude $\hat S_1$ associated with the term with frequency $\nu$ is reduced by
\be
\hat S_1 (\phi) = S_1 \cos (\bar \phi - \phi)  = S_1 \cos \left(\frac{\Psi}{2} + \phi_0 - \phi \right) \ .
\label{eq:demc_nu}
\ee 
For tadpole orbits we have $\Psi \sim 0$ (Fig.\,\ref{fig:modphaT}), which means that $\bar \phi \sim \phi_0$ (Eq.\,\ref{eq:psi0}). Therefore, $\hat S_1$ is maximized for $\phi \sim \phi_0$ and minimized for $\phi \sim \phi_0 + \pi/2$ (Eq.\,\ref{eq:demc_nu}).
Performing a FFT to $\tilde s_k$ allows us to estimate $P_\nu \approx 154.66$~days, $S_1 \approx 4.23$~m/s, and $\Delta \phi \approx -116.5^\circ$. We can also estimate $\bar \phi$ (and hence $\phi_1$ and $\phi_{-1}$) using the ratio between the two amplitudes
\be
\bar \phi = \phi_0 + \arctan \Big( \frac{\hat S_1 (\phi_0 + \pi/2)}{\hat S_1 (\phi_0)} \Big) \approx -4.66^\circ \ .
\ee
Finally, adopting these values as initial parameters, we refit the raw data $s_k$ by performing a minimization of expression (\ref{eq:RV_v2}) using the Levenberg-Marquardt method \citep[e.g.][]{PrNa1992}. The results corresponding to the minimum of $\chi^2$ are shown in Table~\ref{tab:fitspec}.

\begin{table}[h]
\begin{center}
\caption{Fitted parameters using expression (\ref{eq:RV_v2}).}
\label{tab:fitspec}
\begin{footnotesize}
\begin{tabular}{|c|c|c|}
\hline
param. &  tadpole & horseshoe   \\
\hline
\hline
$P_n$\,[day]\dotfill &  $11.4599 {\, \pm \,10^{-4}}$ & $11.5492 {\, \pm 5\,10^{-4}}$ \\ 
$P_\nu$\,[day]\dotfill & $154.66 {\, \pm \, 0.06}$   & $1340 {\, \pm \, 19}$\\  
$\bar S$\,[km/s]\dotfill &  $ 6.5001 {\, \pm \,10^{-4}}$ &  $  6.5001 {\, \pm \,10^{-4}}$ \\
$S_0$\,[m/s]\dotfill &  $ 61.1 {\, \pm \, 0.1}$ &  $ 4.9 {\, \pm \, 0.1}$ \\
$S_1$\,[m/s]\dotfill &  $ 4.23 {\, \pm \, 0.09}$ &  $ 1.2 {\, \pm \, 0.1}$ \\
$S_{-1}$\,[m/s]\dotfill &  $ 4.23 {\, \pm \, 0.09}$ &  $ 1.2 {\, \pm \, 0.1}$ \\
$\phi_0$\,[deg]\dotfill & $341.6 {\, \pm \, 0.1}$ & $22.93 {\, \pm \, 1.69}$   \\
$\phi_1$\,[deg]\dotfill &  $266.4 {\, \pm \, 1.8}$ & $ 309.3 {\, \pm \, 6.4}$\\
$\phi_{-1}$\,[deg]\dotfill &  $ 33.3 {\, \pm \, 2.1}$& $ 280.2 {\pm 7.9}$ \\
\hline
$\sqrt{\chi^2}$ \dotfill & $2.570$ & $1.613$  \\
rms[$m.s^{-1}$]\dotfill & $2.8217$ & $1.8489$ \\
\hline
$A_m$\dotfill & $0.069$ & $0.247$   \\
$\Psi$\,[deg]\dotfill & $-23.5$ & $183.64$   \\
\hline
\end{tabular}
\end{footnotesize}
\end{center}
\end{table} 

\begin{table}[h]
\begin{center}
\caption{Osculating orbital elements obtained through the inversion of the harmonic terms fitted to the observational data (Table~\ref{tab:fitspec}). The elements marked with $^*$ cannot be determined with the Keplerian circular orbit approximation (Eq.\,\ref{eq:stk}), so they have been fixed at constant values. ** indicates that the displayed mass is the lowest possible value ($m \sin I$).}
\label{tab:parinv}
\begin{footnotesize}
\begin{tabular}{|c|cc|cc|}
\hline
 &\multicolumn{2}{c|} {tadpole} &\multicolumn{2}{c|} {horseshoe}  \\ 
param. &  planet 1 & planet 2  &  planet 1 & planet 2   \\
\hline
\hline
$m$\,[$M_{\oplus}$]\dotfill  & $226.4$ & $101.6$ &$18.19$** & $2.74$**  \\
$a$\,[au]\dotfill & 0.099 & 0.101 &  0.1002 & 0.0987 \\
$\lambda$\,[deg]\dotfill & 1.380 & 303.5 &  5.10  & 320.78 \\ 
$e$\dotfill & 0* & 0* & 0* & 0* \\ 
$\varpi$\,[deg]\dotfill & 0* & 0* & 0* & 0*  \\ 
$I$\,[deg]\dotfill & 59.85   & 59.85 & 90*   & 90* \\ 
\hline
$\bar{a}$\,[au]\dotfill &\multicolumn{2}{c|}  {$0.09948$} & \multicolumn{2}{c|} {$0.09999$}  \\
$\zeta_0$\,[deg]\dotfill &\multicolumn{2}{c|}  {$38.01$} & \multicolumn{2}{c|} {18.5*}  \\
$\delta$\,\dotfill &\multicolumn{2}{c|}  {$0.3440$} & \multicolumn{2}{c|} {$0.1309$}  \\
\hline
\end{tabular}
\end{footnotesize}
\end{center}
\end{table}

\begin{table}[h]
\begin{center}
\caption{Best fitted orbital solution using the direct n-body equations of motion, and adopting the orbital parameters listed in Table~\ref{tab:parinv} as the starting point.}
\label{tab:parf}
\begin{footnotesize}
\begin{tabular}{|c|cc|}
\hline
 &\multicolumn{2}{c|} {tadpole}  \\ 
param. &  planet 1 & planet 2     \\
\hline
\hline
$m$\,[$M_{\oplus}$]\dotfill &  $195.68 \pm 0.31$ &  $100.40 \pm 0.35$  \\
$a$\,[au]\dotfill &   0.099 $\pm 6\,10^{-5}$ & 0.101 $\pm 1\,10^{-4}$\\
$\lambda$\,[deg]\dotfill &  $2.3 {\pm 1.7}$ & $306 {\pm 5}$ \\ 
$e$\dotfill & $0.056 {\pm 0.003}$ & $0.049 {\pm 0.003}$  \\ 
$\varpi$\,[deg]\dotfill &  $0.01 {\pm 0.04}$ & $304.2 {\pm 0.4}$ \\ 
$I$\,[deg]\dotfill  & $65 {\pm 2}$ & $57 {\pm 3}$ \\
\hline
$\bar{a}$\,[au]\dotfill &\multicolumn{2}{c|}  {$0.09953$}  \\
$\zeta_0$\,[deg]\dotfill &\multicolumn{2}{c|}  {$37.40$}   \\
$\delta$\,\dotfill &\multicolumn{2}{c|}  {$0.3391$}   \\
\hline
$\sqrt{\chi^2}$ \dotfill & \multicolumn{2}{c|} {$1.652$}  \\
rms[$m.s^{-1}$]\dotfill & \multicolumn{2}{c|} {$1.8770$}  \\  
\hline
\end{tabular}
\begin{tabular}{|c|cc|}
\hline
 &\multicolumn{2}{c|} {horseshoe}  \\ 
param.  &  planet 1 & planet 2   \\
\hline
\hline
$m$\,[$M_{\oplus}$]\dotfill & $18.79 \pm 0.008$ & $2.99 \pm 0.005$  \\
$a$\,[au]\dotfill &     0.100 $\pm 2\,10^{-5}$ & 0.099 $\pm 7\,10^{-5}$ \\
$\lambda$\,[deg]\dotfill &  $4.69 {\pm 2.83}$ & $318.42 {\pm 8.56}$ \\ 
$e$\dotfill &   $0.000 {\pm 10^{-3}}$ & $0.000 {\pm 10^{-3}}$ \\ 
$\varpi$\,[deg]\dotfill &   $0.000 {\pm 10^{-3}}$ & $0.000 {\pm 10^{-3}}$ \\ 
$I$\,[deg]\dotfill  &  90*  & 90* \\
\hline
$\bar{a}$\,[au]\dotfill & \multicolumn{2}{c|} {$0.1000$}  \\
$\zeta_0$\,[deg]\dotfill &\multicolumn{2}{c|} {$21.57$}  \\
$\delta$\,\dotfill & \multicolumn{2}{c|} {$0.1371$}  \\
\hline
$\sqrt{\chi^2}$ \dotfill & \multicolumn{2}{c|} {$1.595$}  \\
rms[$m.s^{-1}$]\dotfill & \multicolumn{2}{c|} {$1.8513$} \\  
\hline
\end{tabular}
\end{footnotesize}
\end{center}
\end{table}   


From the observational parameters listed in Table \ref{tab:fitspec}, we can obtain the corresponding orbital parameters using the inversion method explained in section \ref{sss:inv}. The osculating orbital elements are then obtained through the equations (\ref{eq:lamb}) and (\ref{eq:sol_orb_zeta}). The results are given Table \ref{tab:parinv}.
Except for the eccentricities and the longitudes of the pericentre, which cannot be determined with a Keplerian circular orbit approximation (Eq.\,\ref{eq:stk}),
we obtain a very good agreement for the remaining parameters (cf. Table\,\ref{tab:pari}). 

We can still improve the quality of the fit in a last step, by performing an adjustment to the data using the direct n-body equations of motion \citep[e.g.][]{CoCo2010}. By adopting the orbital parameters listed in Table~\ref{tab:parinv} as the starting point, the algorithm converges rapidly to the best fit. The results are given in Table~\ref{tab:parf}. This last step slightly improves the orbital parameters obtained previously (lower $\chi^2$ and rms), because it is able to additionally fit the eccentricities and the longitudes of the pericentre.
We note, however, that the n-body algorithm is only able to converge to the correct orbital solution because it used the parameters from Table~\ref{tab:parinv} as starting point.
Indeed, the phase space of co-orbital planets has many other local minima that provide alternative solutions that are not real.

%

\subsubsection{Horseshoe orbits}

\begin{figure}[h!]
\begin{center}
\vspace{.5cm}
\hspace{5cm} (a) \hspace{7cm} (b) \vspace{7cm}\\
\hspace{5cm} (c) \hspace{7cm} (d) \vspace{-8.7cm}

 \includegraphics[width=.49\linewidth]{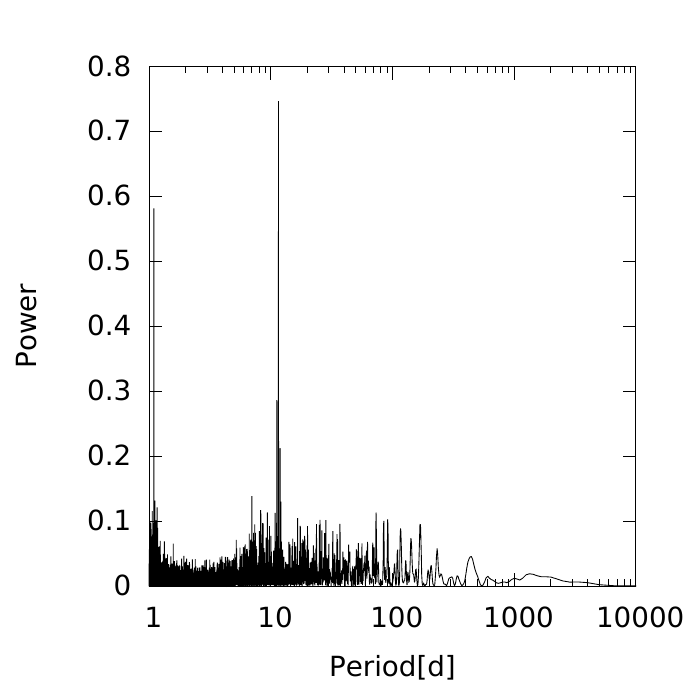}
 \includegraphics[width=.49\linewidth]{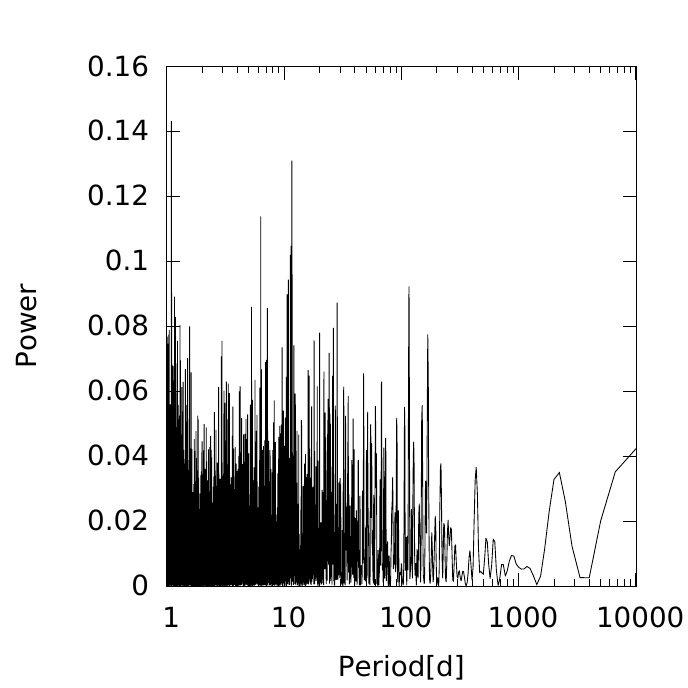} \\
\includegraphics[width=.49\linewidth]{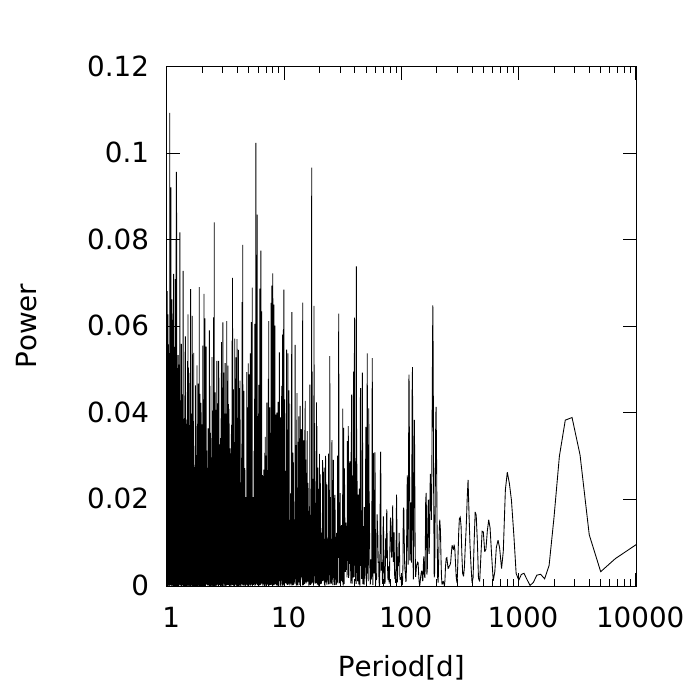}
\includegraphics[width=.49\linewidth]{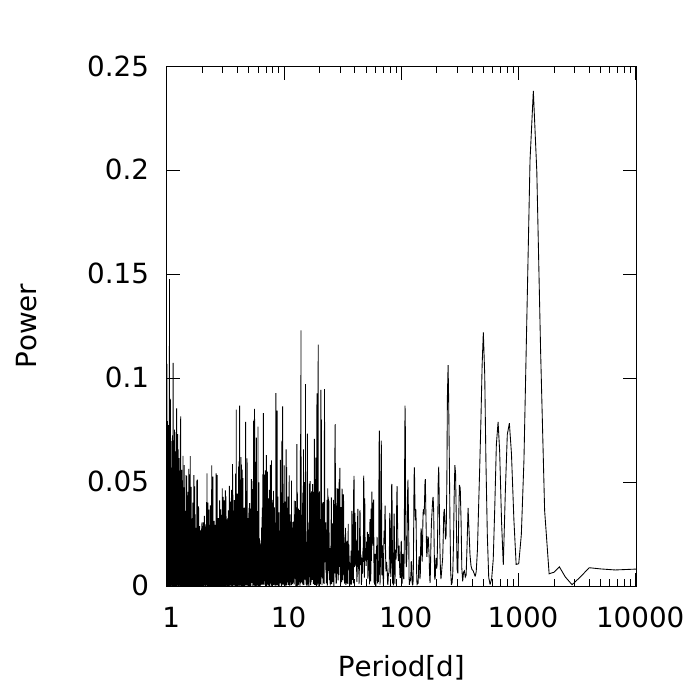}
\caption{Periodograms of the synthetic radial velocity of the horseshoe configuration presented in Table \ref{tab:pari}. \textit{(a)} raw data $s_k$; \textit{(b)} modified data $s_k'$, after the subtraction the Keplerian signal (Eq.\,\ref{eq:demkp}); \textit{(c)} modified data $\tilde s_k$ with $\phi = \phi_0$ (Eq.\,\ref{eq:dem}); and \textit{(d)} modified data $\tilde s_k$ with $\phi = \phi_0 + \pi/2$ (Eq.\,\ref{eq:dem}).}
\label{fig:fft_hs}
\end{center} 
\end{figure}

Our horseshoe system is composed of a Neptune-mass and a 3 Earth-mass planet at $0.1$~au. It is at the limit of detection, since the individual RV amplitudes of each planet are $K = 4.85$~m/s and $K = 0.85$~m/s, respectively.
With this example we intend to show the limitations of our method. 

In Figure~\ref{fig:fft_hs}\,(a), we show a generalized Lomb-Scargle periodogram of the RV data. As for the tadpole example in the  previous section (Fig.\,\ref{fig:fft_t}), the Keplerian component of the signal can clearly be identified for a period $P_n \approx 11 $~days. 
We thus fit the raw data with a Keplerian function (Eq.\,\ref{eq:stk}) obtaining an initial estimation for $P_n \approx 11.55$~days, $\bar S \approx 6.5$~km/s, $S_0 \approx 4.9$~m/s, and $\phi_0 \approx 23^\circ$, subtract its contribution to the data, and obtain a modified data set $s_k'$ (Eq.\,\ref{eq:demkp}).
However, unlike  the tadpole case, in the new periodogram of the residual data, there is no clear peak above the noise (Fig.\,\ref{fig:fft_hs}\,b). Therefore, such a system can easily be mistaken with a system hosting a single planet at 11 days.

We can nevertheless apply our method to search for the traces of a co-orbital companion.
We thus modify the data $\tilde s_k$ according to expression (\ref{eq:dem}) adopting $\phi = \phi_0 = 23^\circ$ and $\phi = \phi_0 + \pi/2 = 113^\circ$.
In Figures~\ref{fig:fft_hs}\,(c) and \ref{fig:fft_hs}\,(d) we show the periodograms corresponding to these transformations, respectively.
For $\phi = \phi_0$ the periodogram is very similar to the one with the residual data (Fig.\,\ref{fig:fft_hs}\,b), so we conclude there is nothing else above the noise in the data.
However, for $\phi = \phi_0 + \pi/2$ the scenario is completely different as a significant peak appears around 1500 days, corresponding to the libration frequency (Fig.\,\ref{fig:fft_hs}\,d).
Indeed, for horseshoe orbits we have $\Psi = \pi$ (Eq.\,\ref{eq:PmHS}), which means that $\bar \phi = \phi_0 + \pi/2$ (Eq.\,\ref{eq:psi0}). Therefore, $\hat S_1$ is null for $\phi = \phi_0$ and maximized for $\phi = \phi_0 + \pi/2$ (Eq.\,\ref{eq:demc_nu}).

Performing a FFT to $\tilde s_k$ allow us to estimate $P_\nu \approx 1340$~days, $S_1 \approx 1.2$~m/s, and $\Delta \phi \approx 295^\circ$. Adopting these values as initial parameters, we refit the raw data $s_k$ with expression (\ref{eq:RV_v2}). The results corresponding to the minimum of $\chi^2$ are shown in Table~\ref{tab:fitspec}.
Comparing these results to the tadpole case, we observe that the uncertainty associated with the $S_p$ and $\phi_p$ terms is larger, but still near 1~m/s, which corresponds to the considered precision of the instrument.
Our method is therefore able to extract any information on the existence of a co-orbital companion, provided that the information on the libration terms is accessible in the data.

Once the existence of a co-orbital companion is confirmed, we can determine the corresponding orbital parameters. The parameter $\zeta_0$ (which gives the departure of the semi-major axis and the mean longitudes from their mean value) has a low impact on the orbital parameters and cannot be easily determined in horseshoe configuration (see section~\ref{sss:invhs}). However, its value is constrained by the stability of the system: in the horseshoe configuration, it ranges between its lowest stable value for $\mu_{min}=\mu\,\sin I$, in our case $\approx 6 \times 10^{-5}$ (see Fig.\,\ref{fig:stabzr}), and the separatrix. We therefore have $\zeta_0 \in [13^\circ, 24^\circ]$. We take $\zeta_0=18.5^\circ$ (average value on this interval) and compute the corresponding orbital parameters. We obtain a system close to the original one (Tab.\,\ref{tab:parinv}).

In the horseshoe case, we cannot  determine either the eccentricities and the longitudes of the pericentre, because we used a Keplerian circular orbit approximation (Eq.\,\ref{eq:stk}),
or the inclination to the line of sight, because we only fit the first three harmonics (Eq.\,\ref{eq:RV_v2}).
In a final step, we perform an adjustment to the data using the direct n-body equations of motion, and we obtain a similar adjustment (Tab.\,\ref{tab:parf}).


\section{Discussion and conclusion}  

In this paper we have revisited the dynamics of quasi-circular co-orbital planets. By computing their gravitational effect on the parent star, we have found a simple method for detecting this kind of planets, provided that the orbital libration can be seen in the observational data. Indeed, when the star is accompanied by co-orbital planets, in addition to the Keplerian orbital motion, there is a modulation at a longer period, corresponding to the libration frequency. Therefore, commonly used methods for signal demodulation (see section \ref{sec:dem}) can also be applied to co-orbital systems, allowing   the amplitude and the frequency of the modulation to be identified more accurately.

Every time a modulation is observed in the motion of a single planet, an inquiry should be made to check if it can correspond to the libration induced by another co-orbital planet.
In this paper, we explain a way to quantify which co-orbital configurations can be expected:
for stability reasons, we can put boundaries for pairs of the parameters ($\mu$, $\zeta_0$); 
for data span duration reasons, we can estimate the frequency of libration $\nu$, depending on ($n$, $\mu$, $\zeta_0$); 
for measurement precision reasons, we can estimate the amplitude of the modulating peaks, which depends on the parameters ($\mu$, $n$, $\delta$, $\zeta_0$).

For reasons of clarity, we exemplify our method in the case of a radial velocity signal. However, our results are valid for any other method that measures a projection of the stellar motion. We have shown that the relative amplitude of the modulation signal depends only on the distance to the Lagrangian equilibrium, $\zeta_0$, and mass ratio, $\delta$. Therefore, the detection of co-orbital planets is enhanced for large libration amplitudes around the Lagrangian equilibrium (i.e. small $\zeta_0$ values), and for planetary masses equally distributed between the two co-orbitals ($\delta \approx 1/2$).

In order to reduce the data, we proposed a direct inversion from the periodograms of the signal to the osculating elements of the system. 
For systems in the tadpole configuration we are able to determine the inclination of the orbital plane with respect to the plane of the sky and hence the true masses of the planets (and not only the minimum masses). In the horseshoe case this is not possible without considering higher harmonics for the modulation.\\


\vspace{1cm}

acknowledgements: \textit{ We acknowledge support from CIDMA strategic project UID/MAT/04106/2013.
The ``conseil scientifique de l'Observatoire de Paris'' is  acknowledged for their financial support. }

\


%
%
%
%
%
%
%
%
%


\appendix
\section{Symmetries}

Equation (\ref{eq:sol_orb_zeta}), respectively (\ref{eq:sol_orb_zeta_tau}), possesses several symmetries. We use two of them to study analytically some features of the horseshoe configuration. On the one hand, we have the symmetry with respect to $\dot{\zeta}=0$ ($\Delta a/a=0$ in Fig.\,\ref{fig:orb}):
\be
\zeta(-(t-t_0))=\zeta(t-t_0)\ .
\label{eq:sym0}
\ee
On the other hand, we have the central symmetry of the phase space ($\zeta,\Delta a/a$) in $\zeta=\pi$ and $\Delta a/a=0$:
\be
\zeta\left(t-t_0 +\frac{\pi}{\nu} \right)=2\pi- \zeta(t-t_0)\ .
\label{eq:sym2}
\ee
Similar expressions can be obtained for $\hat{\zeta}$. In the tadpole configuration, these symmetries exist as well, but the symmetry (\ref{eq:sym2}) maps a vicinity of $L_4$ to a vicinity of $L_5$. 

We can use these symmetries to simplify the expression of the coefficients $C_p$ given by equation (\ref{eq:Cp}). Our purpose is to study the values of $A_m(\delta,\zeta_0)$ and $\Psi(\delta,\zeta_0)$. Since none of them depends on the value of $\tau_0$, we take $\tau_0=0$ from now on. The coefficients $c_p$ (Eq.~\ref{eq:cp_v1}) become
\be
c_p=c_p(\delta,\zeta_0,0)=\frac{\tilde{\nu}}{2\pi}\int^{\pi/\tilde{\nu}}_{-\pi/\tilde{\nu}}e^{i\delta\tilde{\zeta}(\tau)} e^{-ip\tilde{\nu} \tau} d\tau \ .
\label{eq:cp_v1a}
\ee 
Since we took $\tau_0=0$, $e^{-i\delta \zeta}$ is an even function in the case of a horseshoe orbit. Hence $e^{-ip\tilde{\nu}\tau}$ becomes $\cos(p\tilde{\nu}\tau)$ in the expressions of the $C_p$.  By splitting this expression into two integrals and changing $\tau$ to $\tau + \pi/\tilde{\nu}$ in the first one, we get
\be
c_p=\frac{\tilde{\nu}}{2\pi} \int^{\frac{\pi}{\tilde{\nu}}}_{0} \left[e^{i\delta \tilde{\zeta}(\tau+\frac{\pi}{\tilde{\nu}})}e^{-ip\pi}+e^{i\delta\tilde{\zeta}(\tau)} \right] \cos(p\tilde{\nu}\tau) d\tau \ .
\label{eq:cp_v2}
\ee 
Then, using the symmetry given by expression (\ref{eq:sym2}), the previous integral simplifies as
\be
c_p=\frac{\tilde{\nu}}{2\pi} \int^{\frac{\pi}{\tilde{\nu}}}_{0} \left[e^{i\delta (2\pi -\tilde{\zeta}(\tau))}e^{-ip\pi}+e^{i\delta\tilde{\zeta}(\tau)} \right] \cos(p\tilde{\nu}\tau) d\tau ,
\label{eq:cp_v3}
\ee
hence
%
\be
c_p=\frac{\tilde{\nu}}{\pi} e^{i\pi(\delta-\frac{p}{2})} \int^{\frac{\pi}{\tilde{\nu}}}_{0}  \cos (\delta(\pi-\tilde{\zeta}(\tau))-\frac{p\pi}{2}) \cos(p\tilde{\nu}\tau) d\tau .
\label{eq:cp_v4}
\ee
As a consequence, using equation (\ref{eq:Cp}), we get for $p=0$
%
%
\be
\begin{split}
C_0= \frac{\tilde{\nu}}{\pi} e^{i\delta\pi} \int^{\frac{\pi}{\tilde{\nu}}}_{0} \left[ (1-\delta)\cos (\delta(\pi-\tilde{\zeta}(\tau))) \right.  \\
\left. - \delta \cos ((\delta-1)(\pi-\tilde{\zeta}(\tau))) \right] d\tau \ ,
\end{split}
\label{eq:C0_v2}
\ee
and for $q =\pm 1$
\be
\begin{split}
C_{q}= \frac{\tilde{\nu}}{\pi} e^{i(\delta\pi-\frac{\pi}{2})} \int^{\frac{\pi}{\tilde{\nu}}}_{0} \left[ (1-\delta)\sin (\delta(\pi-\tilde{\zeta}(\tau))) \right.  \\
\left. - \delta \sin ((\delta-1)(\pi-\tilde{\zeta}(\tau))) \right] \cos(q\tilde{\nu}\tau) d\tau \ .
\end{split}
\label{eq:C1_v1}
\ee
 We obtain $C_{1} = C_{-1}$.

\subsection{Computation of $\Psi$}
\label{ap:psi}
From equation (\ref{eq:C0_v2}), we have $arg(C_0(\delta))=\delta\pi$ if $\delta \in [0,1/2[$ and $\delta\pi+\pi$ if $\delta \in ]1/2,1]$. Since $arg(C_1)=arg(C_{-1})=(\delta-1/2)\pi$ (Eq.~\ref{eq:C1_v1}), we conclude that for any horseshoe configuration
 \be
\Psi= arg(C_1)+arg (C_{-1}) - 2arg(C_0)  = \pi \ ,
\label{eq:PmHS}
\ee
i.e. $\Psi$ is constant and equal to $\pi$. 

\subsection{Computation of $A_m$}
\label{ap:am}
Generally, the $|C_q|$ (Eq.~\ref{eq:C1_v1}) does not have an explicit expression. However, $A_m$ can be computed for some specific values of $\delta$.
 We denote $C^\delta_q=C_q(\delta,\zeta_0,0)$. For $\delta=1/2$, we have
\be
\begin{split}
C_{q}^{1/2}= \frac{\tilde{\nu}}{\pi} \int^{\frac{\pi}{\tilde{\nu}}}_{0} \left[ \sin ((\pi-\tilde{\zeta}(\tau))/2) \right] \cos(q\tilde{\nu}\tau) d\tau \ .
\end{split}
\label{eq:C1_v3}
\ee
The amplitude of the first harmonics ($q =\pm 1$) of the Fourier series of an odd function is not null. Thus, since from expression (\ref{eq:C0_v2}) we have that $|C_0^{1/2}|=0$, we can conclude that  in the horseshoe configuration $A_m(\frac{1}{2},\zeta_0) = \infty $ (Eq.\,\ref{eq:Am}). Similarly, one can also see from equations (\ref{eq:C0_v2}) and (\ref{eq:C1_v1}) that $A_m(0,\zeta_0)=A_m(1,\zeta_0)=0$.

\section{Mass ratios}

In section \ref{dnLe}, we have shown that in the vicinity of the Lagrangian equilibrium, a planet with mass $m_1$ is easier to identify when its co-orbital companion is much more massive ($m_1 \ll m_2$) rather than when $m_1 \approx m_2$. We show here that this result holds true in the  horseshoe configuration.
Using the symmetries (\ref{eq:sym0}) and (\ref{eq:sym2}), one can rewrite $C_1$ (Eq.~(\ref{eq:C1_v1})) as
\be
\begin{split}
C_{1}= \frac{2\tilde{\nu}}{\pi} e^{i(\delta\pi-\frac{\pi}{2})} \int^{\frac{\pi}{2\tilde{\nu}}}_{0} \left[ (1-\delta)\sin (\delta(\pi-\tilde{\zeta}(\frac{\pi}{2\tilde{\nu}}-\tau))) \right.  \\
\left. - \delta \sin ((\delta-1)(\pi-\tilde{\zeta}(\frac{\pi}{2\tilde{\nu}}-\tau))) \right] \sin(\tilde{\nu}\tau) d\tau.
\end{split}
\label{eq:C1_v3}
\ee
For a mass $m_1$, we want to compare the quantity $\vS_1= A_m \vS_0 = \alpha|C_1|$ in the case of $m_1=m_2$ ($\delta=1/2$) against the case when $m_1 \ll m_2$ ($\delta \approx 1-m_1/m_2 = 1-\epsilon$). Writing $X(\tau)=\pi-\tilde{\zeta}(\frac{\pi}{2\tilde{\nu}}-\tau)$, from equation (\ref{eq:C1_v1}) we have
\be
\begin{split}
|C_{1}^{1/2}|= \left|\frac{2\tilde{\nu}}{\pi}  \int^{\frac{\pi}{2\tilde{\nu}}}_{0} \left[ \sin (X(\tau)/2)  \right]  \sin(\tilde{\nu}\tau) d\tau \right|,
\end{split}
\label{eq:C1_1s2}
\ee
and at first order in $\epsilon$, equation (\ref{eq:C1_v1}) yields
\be
\begin{split}
|C_{1}^{1-\epsilon}|= \left|\epsilon \frac{\tilde{2\nu}}{\pi} \int^{\frac{\pi}{2\tilde{\nu}}}_{0} \left[  X(\tau)  + \sin (X(\tau)) \right]  \sin(\tilde{\nu}\tau) d\tau \right|.
\end{split}
\label{eq:C1_eps}
\ee

We have $X(0)=0$ and $X(\frac{\pi}{2\tilde{\nu}})=\pi-\zeta_0$. One can show that $X$ is a monotonous function in the interval $\tau \in [0, \pi/(2\tilde{\nu})]$, hence $\sin(X/2)$ and $X+\sin(X)$ are a positive function in this interval. Moreover, for $X \in [0, \pi]$, we have the following inequality:
 \be
\pi \sin(X/2) \leq (X+\sin(X)) \leq 4 \sin(X/2) \ .
\label{eq:C1_gend}
\ee
Since $\sin(\tilde{\nu}\tau)$ is also a positive function on the considered interval, the inequality in equation (\ref{eq:C1_gend}) holds true when we multiply each term by  $\sin(\tilde{\nu}\tau)$ and integrate over $\tau \in [0, \pi/(2\tilde{\nu})]$. Finally, we get
 \be
\pi C_{1}^{1/2} \leq C_{1}^{1-\epsilon}/\epsilon \leq 4 C_{1}^{1/2}.
\label{eq:C1_gend2}
\ee
When $\delta=1/2$, we have $\mu\approx 2 m_1/m_0$, while when $\delta=1-\epsilon$, we get $\mu\approx m_1/(\epsilon m_0)$. Multiplying equation (\ref{eq:C1_gend2}) by $\alpha$, we obtain
 \be
\frac{\pi}{2}  \vS_1^{1/2} \leq \vS_{1}^{1-\epsilon} \leq 2 \vS_{1}^{1/2}.
\label{eq:C1_gend3}
\ee
We finally conclude that in the horseshoe case, for a given mass $m_1$, the co-orbital couple ($m_1$,$m_2$) is up to two times easier to identify when $m_1 \ll m_2$ rather than when $m_1 \approx m_2$. 

\bibliographystyle{aa}

\end{document}